# Application of Non-local Quantum Hydrodynamics to the Description of the Charged Density Waves in the Graphen Crystal Lattice.

Boris V. Alexeev, Irina V. Ovchinnikova
*Moscow Lomonosov University of Fine Chemical Technologies (MITHT)*
*Prospekt Vernadskogo, 86, Moscow 119570, Russia*
Boris.Vlad.Alexeev@gmail.com
e-mail: *boris.vlad.alexeev@gmail.com*

The motion of the charged particles in graphen in the frame of the quantum non-local hydrodynamic description is considered. It is shown as results of the mathematical modeling that the mentioned motion is realizing in the soliton forms. The dependence of the size and structure of solitons on the different physical parameters is investigated.

1. **Introduction. Preliminary remarks.**

The possibility of the non local physics application in the theory of superconductivity is investigated in [1-3]. It is shown that by the superconducting conditions the relay ("estafette") motion of the soliton' system ("lattice ion – electron") is realizing by the absence of chemical bonds. From the position of the quantum hydrodynamics the problem of creation of the high temperature superconductors leads to finding of materials which lattices could realize the soliton' motion without destruction. These materials should be created using the technology of quantum dots.

Non-local physics demonstrates its high efficiency in many fields – from the atom structure problems to cosmology [4 - 16]. Mentioned works contain not only strict theory, but also delivering the qualitative aspects of theory without excessively cumbersome formulas. As it is shown (see, for example [4,5,7 - 11]) the theory of transport processes (including quantum mechanics) can be considered in the frame of unified theory based on the non-local physical description.

This paper is directed on investigation of possible applications of the non-local quantum hydrodynamics in the theory of transport processes in graphen including the effects of the charge density waves (CDW). Is known that graphene, a single-atom-thick sheet of graphite, is a new material which combines aspects of semiconductors and metals. For example the mobility, a measure of how well a material conducts electricity, is higher than for other known material at room temperature. In graphene, a resistivity is of about 1.0 microOhm-cm (resistivity defined as a specific measure of resistance; the resistance of a piece material is its resistivity times its length and divided by its cross-sectional area). This is about 35 percent less than the resistivity of copper, the lowest resistivity material known at room temperature.

Measurements lead to conclusion that the influence of thermal vibrations on the conduction of electrons in graphene is extraordinarily small. From the other side the typical reasoning exists:

"In any material, the energy associated with the temperature of the material causes the atoms of the material to vibrate in place. As electrons travel through the material, they can bounce off these vibrating atoms, giving rise to electrical resistance. This electrical resistance is "intrinsic" to the material: it cannot be eliminated unless the material is cooled to absolute zero temperature, and hence sets the upper limit to how well a material can conduct electricity."

Obviously this point of view leads to the principal elimination of effects of the high temperature superconductivity. From the mentioned point of view the restrictions in mobilities of known semiconductors can be explained as the influence of the thermal vibration of the atoms. The limit to mobility of electrons in graphen is about 200,000 $cm^2/(V \cdot s)$ at room temperature, compared to about 1,400 $cm^2/(V \cdot s)$ in silicon, and 77,000 $cm^2/(V \cdot s)$ in indium antimonide, the highest mobility conventional semiconductor known. The opinion of a part of investigators can be formulated as follows: "Other extrinsic sources in today's fairly dirty graphene samples add some extra resistivity to graphene," (see for example [17]) "so the overall resistivity isn't quite as low as copper's at room temperature yet. However, graphene has far fewer electrons than copper, so in graphene the electrical current is carried by only a few electrons moving much faster than the electrons in copper." Mobility determines the speed at which an electronic device (for instance, a field-effect transistor, which forms the basis of modern computer chips) can turn on and off. The very high mobility makes graphene promising for applications in which transistors much switch extremely fast, such as in processing extremely high frequency signals. The low resistivity and extremely thin nature of graphene also promises applications in thin, mechanically tough, electrically conducting, transparent films. Such films are sorely needed in a variety of electronics applications from touch screens to photovoltaic cells.

In the last years the direct observation of the atomic structures of superconducting materials (as usual superconducting materials in the cuprate family like $YBa_2Cu_3O_{6.67}$ ($T_c = 67$ K)) was realized with the scanning tunneling microscope (STM) and other instruments, STMs scan a surface in steps smaller than an atom.

Superconductivity, in which an electric current flows with zero resistance, was first discovered in metals cooled very close to absolute zero. New materials called cuprates - copper oxides "doped" with other atoms -- superconduct as "high" as minus 123 Celsius. Some conclusions from direct observations [18, 19]:

1. Observations of high-temperature superconductors show an "energy gap" where electronic states are missing. Sometimes this energy gap appears but the material still does not superconduct -- a so-called "pseudogap" phase. The pseudogap appears at higher temperatures than any superconductivity, offering the promise of someday developing materials that would superconduct at or near room temperature.

2. STM image of a partially doped cuprate superconductor shows regions with an electronic "pseudogap". As doping increases, pseudogap regions spread and connect, making the whole sample a superconductor.

3. High temperature superconductivity in layered cuprates can develop from an electronically ordered state called a charge density wave (CDW). The results of observation can be interpreted as the creation of the "checkerboard pattern" due to the modulation of the atomic positions in the $CuO_2$ layers of $YBa_2Cu_3O_{6+x}$ caused by the charge density wave.

4. Application of the method of high-energy X-ray diffraction shows that a CDW develops at zero field in the normal state of superconducting $YBa_2Cu_3O_{6.67}$ ($T_c = 67$ K). Below $T_c$ the application of a magnetic field suppresses superconductivity and enhances the CDW. It means that the high-$T_c$ superconductivity forms from a pre-existing CDW environment.

*Important conclusion*: high temperature superconductors demonstrate new type of electronic order and modulation of atomic positions. As it was shown in [1,3] the delivered above graphene properties can be explained only in the frame of the self-consistent non-local quantum theory (see for example [4,5]) which leads to appearance of the soliton waves moving in graphene.

**2. Generalized quantum hydrodynamic equations describing the soliton movement in the crystal lattice.**

Let us consider the charge density waves which are periodic modulation of conduction electron density. From direct observations of charge density waves follow that CDW develop at zero external fields. For our aims is sufficient in the following to suppose that the effective charge movement was created in grapheme lattice as result of an initial fluctuation.

The movement of the soliton waves at the presence of the external electrical potential difference will be considered also in this article.

This effective charge is created due to interference of the induced electron waves and correlating potentials as result of the polarized modulation of atomic positions. Therefore in this approach the conduction in grapheme convoys the transfer of the positive ($+e, m_p$) and negative ($-e, m_e$) charges. Let us formulate the problem in detail. The non-stationary 1D motion of the combined soliton is considered under influence of the self-consistent electric forces of the potential and non-potential origin. It was shown [1, 3] that mentioned soliton can exists without a chemical bond formation. For better understanding of the situation let us investigate the situation for the case when the external forces are absent. Introduce the coordinate system ($\xi = x - Ct$) moving along the positive direction of the $x$ axis with the velocity $C = u_0$, which is equal to the phase velocity of this quantum object.

Let us find the soliton type solutions for the system of the generalized quantum equations for two species mixture [1, 3, 5, 11]. The graphene crystal lattice is 2D flat structure which is considered in the moving coordinate system ($\xi = x - u_0 t$, $y$).

Write down the system of equations [1, 3, 5, 11] for the two component mixture of charged particles without taking into account the component's internal energy in the dimensionless form, where dimensionless symbols are marked by tildes. We begin with introduction the scales:

$u = u_0 \tilde{u}$ - hydrodynamic velocity;

$\xi = x_0 \tilde{\xi}$, $y = x_0 \tilde{y}$;

$\varphi = \varphi_0 \tilde{\varphi}$ - self-consistent electric potential;

$\rho_e = \rho_0 \tilde{\rho}_e$, $\rho_p = \rho_0 \tilde{\rho}_p$ - densities for the electron and positive species;

$p_e = \rho_0 V_{0e}^2 \tilde{p}_e$, $p_p = \rho_0 V_{0p}^2 \tilde{p}_p$ - quantum electron pressure and the pressure of positive species, where $V_{0e}$, $V_{0p}$ - the scales for thermal velocities for the electron and positive species;

$F_e = \frac{e\varphi_0}{m_e x_0} \tilde{F}_e$, $F_p = \frac{e\varphi_0}{m_p x_0} \tilde{F}_p$ - the forces acting on the mass unit of electrons and the positive particles, where $m_e$, $m_p$ are masses of electrons and the positive particles.

Non-local parameters can be written in the form (see also [1,3,10,11])

$$\tau_e = \frac{x_0 H}{u_0 \tilde{u}^2}, \quad \tau_p = \frac{m_e x_0 H}{m_p u_0 \tilde{u}^2}, \quad \frac{1}{\tau_{ep}} = \frac{1}{\tau_e} + \frac{1}{\tau_p} = \frac{u_0}{x_0} \frac{\tilde{u}^2}{H}\left(1 + \frac{m_p}{m_e}\right). \quad (1)$$

Dimensionless parameter $H = \frac{N_R \hbar}{m_e x_0 u_0}$ is introduced, $N_R$ - entire number. Let us introduce also the following dimensionless parameters

$$R = \frac{e\rho_0 x_0^2}{m_e \varphi_0}, \quad E = \frac{e\varphi_0}{m_e u_0^2}. \tag{2}$$

Taking into account the introduced values the following system of dimensionless non-local hydrodynamic equations for the 2D soliton description can be written (see also [1 - 5]):

Dimensionless Poisson equation for the self-consistent electric field:

$$\frac{\partial^2 \tilde{\varphi}}{\partial \tilde{\xi}^2} + \frac{\partial^2 \tilde{\varphi}}{\partial \tilde{y}^2} = -4\pi R \left\{ \frac{m_e}{m_p} \left[ \tilde{\rho}_p - \frac{m_e H}{m_p \tilde{u}^2} \frac{\partial}{\partial \tilde{\xi}} (\tilde{\rho}_p (\tilde{u} - 1)) \right] - \left[ \tilde{\rho}_e - \frac{H}{\tilde{u}^2} \frac{\partial}{\partial \tilde{\xi}} (\tilde{\rho}_e (\tilde{u} - 1)) \right] \right\}. \tag{3}$$

Continuity equation for the positive particles:

$$\frac{\partial}{\partial \tilde{\xi}} [\tilde{\rho}_p (1 - \tilde{u})] + \frac{m_e}{m_p} \frac{\partial}{\partial \tilde{\xi}} \left\{ \frac{H}{\tilde{u}^2} \frac{\partial}{\partial \tilde{\xi}} [\tilde{\rho}_p (\tilde{u} - 1)^2] \right\} +$$
$$+ \frac{m_e}{m_p} \frac{\partial}{\partial \tilde{\xi}} \left\{ \frac{H}{\tilde{u}^2} \left[ \frac{V_{0p}^2}{u_0^2} \frac{\partial}{\partial \tilde{\xi}} \tilde{p}_p - \frac{m_e}{m_p} E \tilde{\rho}_p \tilde{F}_{p\xi} \right] \right\} + \frac{m_e}{m_p} \frac{\partial}{\partial \tilde{y}} \left\{ \frac{H}{\tilde{u}^2} \left[ \frac{V_{0p}^2}{u_0^2} \frac{\partial}{\partial \tilde{y}} \tilde{p}_p - \frac{m_e}{m_p} E \tilde{\rho}_p \tilde{F}_{py} \right] \right\} = 0 \tag{4}$$

Continuity equation for electrons:

$$\frac{\partial}{\partial \tilde{\xi}} [\tilde{\rho}_e (1 - \tilde{u})] + \frac{\partial}{\partial \tilde{\xi}} \left\{ \frac{H}{\tilde{u}^2} \frac{\partial}{\partial \tilde{\xi}} [\tilde{\rho}_e (\tilde{u} - 1)^2] \right\} +$$
$$+ \frac{\partial}{\partial \tilde{\xi}} \left\{ \frac{H}{\tilde{u}^2} \left[ \frac{V_{0e}^2}{u_0^2} \frac{\partial}{\partial \tilde{\xi}} \tilde{p}_e - \tilde{\rho}_e E \tilde{F}_{e\xi} \right] \right\} + \frac{\partial}{\partial \tilde{y}} \left\{ \frac{H}{\tilde{u}^2} \left[ \frac{V_{0e}^2}{u_0^2} \frac{\partial}{\partial \tilde{y}} \tilde{p}_e - \tilde{\rho}_e E \tilde{F}_{ey} \right] \right\} = 0 \tag{5}$$

Momentum equation for the *x* direction:

$$\frac{\partial}{\partial \tilde{\xi}} \left\{ (\tilde{\rho}_p + \tilde{\rho}_e) \tilde{u} (\tilde{u} - 1) + \frac{V_{0p}^2}{u_0^2} \tilde{p}_p + \frac{V_{0e}^2}{u_0^2} \tilde{p}_e \right\} - \frac{m_e}{m_p} \tilde{\rho}_p E \tilde{F}_{p\xi} - \tilde{\rho}_e E \tilde{F}_{e\xi} +$$
$$+ \frac{m_e}{m_p} \frac{\partial}{\partial \tilde{\xi}} \left\{ \frac{H}{\tilde{u}^2} \left[ \frac{\partial}{\partial \tilde{\xi}} \left( 2 \frac{V_{0p}^2}{u_0^2} \tilde{p}_p (1 - \tilde{u}) - \tilde{\rho}_p \tilde{u} (1 - \tilde{u})^2 \right) - \frac{m_e}{m_p} \tilde{\rho}_p E \tilde{F}_{p\xi} (1 - \tilde{u}) \right] \right\} +$$
$$+ \frac{\partial}{\partial \tilde{\xi}} \left\{ \frac{H}{\tilde{u}^2} \left[ \frac{\partial}{\partial \tilde{\xi}} \left( 2 \frac{V_{0e}^2}{u_0^2} \tilde{p}_e (1 - \tilde{u}) - \tilde{\rho}_e \tilde{u} (1 - \tilde{u})^2 \right) - \tilde{\rho}_e E \tilde{F}_{e\xi} (1 - \tilde{u}) \right] \right\} +$$
$$+ \frac{H}{\tilde{u}^2} E \left( \frac{m_e}{m_p} \right)^2 \tilde{F}_{p\xi} \left( \frac{\partial}{\partial \tilde{\xi}} (\tilde{\rho}_p (\tilde{u} - 1)) \right) + \frac{H}{\tilde{u}^2} E \tilde{F}_{e\xi} \left( \frac{\partial}{\partial \tilde{\xi}} (\tilde{\rho}_e (\tilde{u} - 1)) \right) -$$
$$- \frac{m_e}{m_p} \frac{\partial}{\partial \tilde{\xi}} \left\{ \frac{H}{\tilde{u}^2} \frac{V_{0p}^2}{u_0^2} \frac{\partial}{\partial \tilde{\xi}} (\tilde{p}_p \tilde{u}) \right\} - \frac{\partial}{\partial \tilde{\xi}} \left\{ \frac{H}{\tilde{u}^2} \frac{V_{0e}^2}{u_0^2} \frac{\partial}{\partial \tilde{\xi}} (\tilde{p}_e \tilde{u}) \right\} -$$

$$-\frac{m_e}{m_p}\frac{\partial}{\partial \tilde{y}}\left\{\frac{H}{\tilde{u}^2}\frac{V_{0p}^2}{u_0^2}\frac{\partial}{\partial \tilde{y}}(\tilde{p}_p\tilde{u})\right\}-\frac{\partial}{\partial \tilde{y}}\left\{\frac{H}{\tilde{u}^2}\frac{V_{0e}^2}{u_0^2}\frac{\partial}{\partial \tilde{y}}(\tilde{p}_e\tilde{u})\right\}+$$

$$+\left(\frac{m_e}{m_p}\right)^2\frac{\partial}{\partial \tilde{\xi}}\left\{\frac{H}{\tilde{u}^2}E[\tilde{F}_{p\xi}\tilde{\rho}_p\tilde{u}]\right\}+\frac{\partial}{\partial \tilde{\xi}}\left\{\frac{H}{\tilde{u}^2}E[\tilde{F}_{e\xi}\tilde{\rho}_e\tilde{u}]\right\}+$$

$$+\left(\frac{m_e}{m_p}\right)^2\frac{\partial}{\partial \tilde{y}}\left\{\frac{H}{\tilde{u}^2}E[\tilde{F}_{py}\tilde{\rho}_p\tilde{u}]\right\}+\frac{\partial}{\partial \tilde{y}}\left\{\frac{H}{\tilde{u}^2}E[\tilde{F}_{ey}\tilde{\rho}_e\tilde{u}]\right\}=0 \qquad (6)$$

Energy equation for the positive particles:

$$\frac{\partial}{\partial \tilde{\xi}}\left[\tilde{\rho}_p\tilde{u}^2(\tilde{u}-1)+5\frac{V_{0p}^2}{u_0^2}\tilde{p}_p\tilde{u}-3\frac{V_{0p}^2}{u_0^2}\tilde{p}_p\right]-2\frac{m_e}{m_p}\tilde{\rho}_p E\tilde{F}_{p\xi}\tilde{u}+$$

$$+\frac{\partial}{\partial \tilde{\xi}}\left\{\frac{H}{\tilde{u}^2}\frac{m_e}{m_p}\left[\begin{array}{l}\frac{\partial}{\partial \tilde{\xi}}\left(-\tilde{\rho}_p\tilde{u}^2(1-\tilde{u})^2+7\frac{V_{0p}^2}{u_0^2}\tilde{p}_p\tilde{u}(1-\tilde{u})+3\frac{V_{0p}^2}{u_0^2}\tilde{p}_p(\tilde{u}-1)-\frac{V_{0p}^2}{u_0^2}\tilde{p}_p\tilde{u}^2-5\frac{V_{0p}^4}{u_0^4}\frac{\tilde{p}_p^2}{\tilde{\rho}_p}\right)-\\ -2\frac{m_e}{m_p}E\tilde{F}_{p\xi}\tilde{\rho}_p\tilde{u}(1-\tilde{u})+\frac{m_e}{m_p}\tilde{\rho}_p\tilde{u}^2 E\tilde{F}_{p\xi}+5\frac{m_e}{m_p}\frac{V_{0p}^2}{u_0^2}\tilde{p}_p E\tilde{F}_{p\xi}\end{array}\right]\right\}-$$

$$-\frac{\partial}{\partial \tilde{y}}\left\{\frac{H}{\tilde{u}^2}\frac{m_e}{m_p}\left[\frac{\partial}{\partial \tilde{y}}\left(\frac{V_{0p}^2}{u_0^2}\tilde{p}_p\tilde{u}^2+5\frac{V_{0p}^4}{u_0^4}\frac{\tilde{p}_p^2}{\tilde{\rho}_p}\right)-\frac{m_e}{m_p}\tilde{\rho}_p E\tilde{F}_{py}\tilde{u}^2-5\frac{m_e}{m_p}\frac{V_{0p}^2}{u_0^2}\tilde{p}_p E\tilde{F}_{py}\right]\right\}-$$

$$-2\frac{H}{\tilde{u}^2}\left(\frac{m_e}{m_p}\right)^2 E\tilde{F}_{p\xi}\left[\frac{\partial}{\partial \tilde{\xi}}(\tilde{\rho}_p\tilde{u}(1-\tilde{u}))\right]-2\frac{H}{\tilde{u}^2}\left(\frac{m_e}{m_p}\right)^3\tilde{\rho}_p E^2\left[(\tilde{F}_{p\xi})^2+(\tilde{F}_{py})^2\right]+$$

$$+2\frac{H}{\tilde{u}^2}\left(\frac{m_e}{m_p}\right)^2 E\tilde{F}_{p\xi}\left[\frac{V_{0p}^2}{u_0^2}\frac{\partial}{\partial \tilde{\xi}}\tilde{p}_p\right]+2\frac{H}{\tilde{u}^2}\left(\frac{m_e}{m_p}\right)^2 E\tilde{F}_{py}\left[\frac{V_{0p}^2}{u_0^2}\frac{\partial}{\partial \tilde{y}}\tilde{p}_p\right]=-\frac{\tilde{u}^2}{Hu_0^2}\left(V_{0p}^2\tilde{p}_p-\tilde{p}_e V_{0e}^2\right)\left(1+\frac{m_p}{m_e}\right)$$

$$(7)$$

Energy equation for electrons:

$$\frac{\partial}{\partial \tilde{\xi}}\left[\tilde{\rho}_e\tilde{u}^2(\tilde{u}-1)+5\frac{V_{0e}^2}{u_0^2}\tilde{p}_e\tilde{u}-3\frac{V_{0e}^2}{u_0^2}\tilde{p}_e\right]-2\tilde{\rho}_e E\tilde{F}_{e\xi}\tilde{u}+$$

$$+\frac{\partial}{\partial \tilde{\xi}}\left\{\frac{H}{\tilde{u}^2}\left[\begin{array}{l}\frac{\partial}{\partial \tilde{\xi}}\left(-\tilde{\rho}_e\tilde{u}^2(1-\tilde{u})^2+7\frac{V_{0e}^2}{u_0^2}\tilde{p}_e\tilde{u}(1-\tilde{u})+3\frac{V_{0e}^2}{u_0^2}\tilde{p}_e(\tilde{u}-1)-\frac{V_{0e}^2}{u_0^2}\tilde{p}_e\tilde{u}^2-5\frac{V_{0e}^4}{u_0^4}\frac{\tilde{p}_e^2}{\tilde{\rho}_e}\right)-\\ -2E\tilde{F}_{e\xi}\tilde{\rho}_e\tilde{u}(1-\tilde{u})+\tilde{\rho}_e\tilde{u}^2 E\tilde{F}_{e\xi}+5\frac{V_{0e}^2}{u_0^2}\tilde{p}_e E\tilde{F}_{e\xi}\end{array}\right]\right\}+$$

$$-\frac{\partial}{\partial \tilde{y}}\left\{\frac{H}{\tilde{u}^2}\left[\frac{\partial}{\partial \tilde{y}}\left(\frac{V_{0e}^2}{u_0^2}\tilde{p}_e\tilde{u}^2+5\frac{V_{0e}^4}{u_0^4}\frac{\tilde{p}_e^2}{\tilde{\rho}_e}\right)-\tilde{\rho}_e E\tilde{F}_{ey}\tilde{u}^2-5\frac{V_{0e}^2}{u_0^2}\tilde{p}_e E\tilde{F}_{ey}\right]\right\}-$$

$$-2\frac{H}{\tilde{u}^2}E\tilde{F}_{e\xi}\left[\frac{\partial}{\partial \tilde{\xi}}(\tilde{\rho}_e\tilde{u}(1-\tilde{u}))\right]-2\frac{H}{\tilde{u}^2}\tilde{\rho}_e E^2\left[(\tilde{F}_{e\xi})^2+(\tilde{F}_{ey})^2\right]+$$

$$+2\frac{H}{\tilde{u}^2}E\tilde{F}_{e\xi}\left[\frac{V_{0e}^2}{u_0^2}\frac{\partial}{\partial \tilde{\xi}}\tilde{p}_e\right]+2\frac{H}{\tilde{u}^2}E\tilde{F}_{ey}\left[\frac{V_{0e}^2}{u_0^2}\frac{\partial}{\partial \tilde{y}}\tilde{p}_e\right]=-\frac{\tilde{u}^2}{Hu_0^2}\left(V_{0e}^2\tilde{p}_e-V_{0p}^2\tilde{p}_p\right)\left(1+\frac{m_p}{m_e}\right)$$

$$(8)$$

The right hand sides of the energy equations are written in the relaxation forms following from BGK kinetic approximation.

Acting forces are the sum of three terms: the self-consistent potential force (scalar potential $\varphi$), connected with the displacement of positive and negative charges, potential forces originated by the grapheme crystal lattice (potential $U$) and the external electrical field creating the intensity **E**. As result the following dimensionless relations are valid

$$\tilde{F}_{p\xi} = -\frac{\partial \tilde{\varphi}}{\partial \tilde{\xi}} - \frac{\partial \tilde{U}}{\partial \tilde{\xi}} + \tilde{E}_\xi, \quad \tilde{F}_{e\xi} = \frac{\partial \tilde{\varphi}}{\partial \tilde{\xi}} + \frac{\partial \tilde{U}}{\partial \tilde{\xi}} - \tilde{E}_\xi,$$

$$\tilde{F}_{py} = -\frac{\partial \tilde{\varphi}}{\partial \tilde{y}} - \frac{\partial \tilde{U}}{\partial \tilde{y}} + \tilde{E}_y, \quad \tilde{F}_{ey} = \frac{\partial \tilde{\varphi}}{\partial \tilde{y}} + \frac{\partial \tilde{U}}{\partial \tilde{y}} - \tilde{E}_y. \tag{9}$$

Graphene is a single layer of carbon atoms densely packed in a honeycomb lattice. Figure 1 reflects the structure of grapheme as the 2D hexagonal carbon crystal, the distance $a$ between the nearest atoms is equal to $a = 0.142\ nm$.

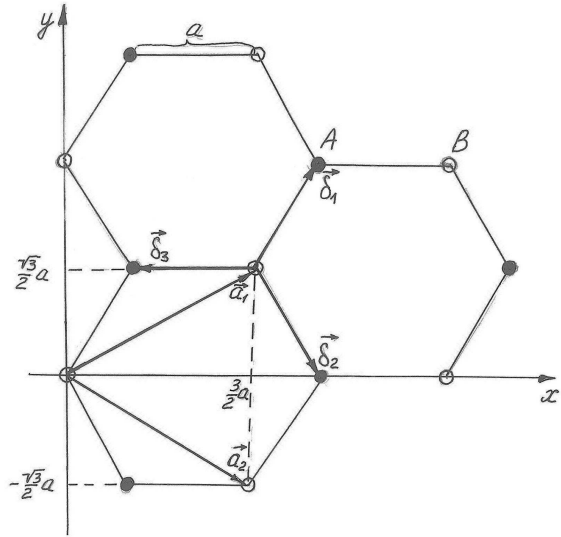

Figure 1. Crystal graphene lattice.

Elementary cell contains two atoms (for example A and B, Fig. 1) and the primitive lattice vectors are given by

$$\mathbf{a}_1 = \frac{a}{2}\left(3; \sqrt{3}\right), \quad \mathbf{a}_2 = \frac{a}{2}\left(3; -\sqrt{3}\right).$$

Coordinates of the nearest atoms to the given atom define by vectors

$$\boldsymbol{\delta}_1 = \frac{a}{2}\left(1; \sqrt{3}\right), \quad \boldsymbol{\delta}_2 = \frac{a}{2}\left(1; -\sqrt{3}\right), \quad \boldsymbol{\delta}_3 = -a(1;0).$$

Six neighboring atoms of the second order are placed in knots defined by vectors

$$\boldsymbol{\delta}'_1 = \pm \mathbf{a}_1, \quad \boldsymbol{\delta}'_2 = \pm \mathbf{a}_2, \quad \boldsymbol{\delta}'_3 = \pm(\mathbf{a}_2 - \mathbf{a}_1).$$

Let us take the first atom of the elementary cell in the origin of the coordinate system (Fig. 1) and compose the radii-vector of the second atom with respect to the basis $\mathbf{a}_1$ и $\mathbf{a}_2$:

$$\mathbf{r}_1 = u\mathbf{a}_1 + v\mathbf{a}_2 = u\left(3\frac{a}{2}\mathbf{e}_x + \sqrt{3}\frac{a}{2}\mathbf{e}_y\right) + v\left(3\frac{a}{2}\mathbf{e}_x - \sqrt{3}\frac{a}{2}\mathbf{e}_y\right). \tag{10}$$

Let us find $u$ и $v$, taking into account that

$$\mathbf{r}_1 = \boldsymbol{\delta}_1 = \frac{a}{2}\left(1;\sqrt{3}\right) = \frac{a}{2}\mathbf{e}_x + \frac{a}{2}\sqrt{3}\mathbf{e}_y. \tag{11}$$

Equalizing (10) и (11), we have $u = \frac{2}{3}$, $v = -\frac{1}{3}$, then

$$\mathbf{r}_1 = \frac{2}{3}\mathbf{a}_1 - \frac{1}{3}\mathbf{a}_2. \tag{12}$$

Assume that $V_1(\mathbf{r})$ is the periodical potential created by one sublattice. Then potential of crystal is

$$V(\mathbf{r}) = V_1(\mathbf{r}) + V_1(\mathbf{r} - \mathbf{r}_1) = \sum_{n=0}^{1} V_1(\mathbf{r} - \mathbf{r}_n). \tag{13}$$

Atoms in crystal form the periodic structure and as the consequence the corresponding potential is periodic function

$$V_1(\mathbf{r}) = V_1(\mathbf{r} + \mathbf{a}_m),$$

where for 2D structure

$$\mathbf{a}_m = m_1\mathbf{a}_1 + m_2\mathbf{a}_2,$$

and $m_1$ и $m_2$ are arbitrary entire numbers. Expanding $V_1(\mathbf{r})$ in the Fourier series one obtains

$$V_1(\mathbf{r} - \mathbf{r}_n) = \sum_{\mathbf{b}} V_{\mathbf{b}} e^{i\mathbf{b}\cdot(\mathbf{r}-\mathbf{r}_n)}. \tag{14}$$

In our case the both basis atoms ($n=0,1$) are the same. Here

$$\mathbf{b} = g_1\mathbf{b}_1 + g_2\mathbf{b}_2,$$

$\mathbf{b}_1$ и $\mathbf{b}_2$ are the translational vectors of the reciprocal lattice. For graphene

$$\mathbf{b}_1 = \frac{2\pi}{3a}\left(1;\sqrt{3}\right), \quad \mathbf{b}_2 = \frac{2\pi}{3a}\left(1;-\sqrt{3}\right). \tag{15}$$

Then

$$V(\mathbf{r}) = \sum_{\mathbf{b}} \sum_{n=0}^{1} V_{1\mathbf{b}} e^{i\mathbf{b}\cdot(\mathbf{r}-\mathbf{r}_n)} = \sum_{\mathbf{b}} V_{\mathbf{b}} e^{i\mathbf{b}\cdot\mathbf{r}}, \tag{16}$$

where $V_{\mathbf{b}} = V_{1\mathbf{b}} \cdot \sum_n e^{-i\mathbf{b}\cdot\mathbf{r}_n} = V_{1\mathbf{b}} \cdot S_{\mathbf{b}}$. The structure factor $S_{\mathbf{b}}$ for graphene:

$$S_{\mathbf{b}} = e^{-i\mathbf{b}\cdot 0} + e^{-i\mathbf{b}\cdot\left(\frac{2}{3}\mathbf{a}_1 - \frac{1}{3}\mathbf{a}_2\right)} = 1 + e^{i\frac{2\pi}{3}(g_2 - 2g_1)}. \tag{17}$$

$$V(\mathbf{r}) = \sum_{g_1, g_2} V_{1g_1, g_2} e^{i(g_1 \mathbf{b}_1 + g_2 \mathbf{b}_2) \cdot \mathbf{r}} \left(1 + e^{i\frac{2\pi}{3}(g_2 - 2g_1)}\right). \tag{18}$$

For the approximate calculation we use the terms of the series with $|g_1| \leq 2$, $|g_2| \leq 2$. Therefore

$$V(\mathbf{r}) = 2V_{1,(00)} +$$
$$+ 4V_{1,(10)} \left( \cos\left(\frac{1}{2}(\mathbf{b}_1 + \mathbf{b}_2) \cdot \mathbf{r}\right) \cos\left(\frac{1}{2}(\mathbf{b}_1 - \mathbf{b}_2) \cdot \mathbf{r}\right) + \cos\left(\frac{1}{2}(\mathbf{b}_1 + \mathbf{b}_2) \cdot \mathbf{r} + \frac{2\pi}{3}\right) \cos\left(\frac{1}{2}(\mathbf{b}_1 - \mathbf{b}_2) \cdot \mathbf{r}\right) \right) +$$
$$+ 2V_{1,(11)} \left( \cos((\mathbf{b}_1 + \mathbf{b}_2) \cdot \mathbf{r}) + \cos\left((\mathbf{b}_1 + \mathbf{b}_2) \cdot \mathbf{r} - \frac{2\pi}{3}\right) + 2\cos((\mathbf{b}_1 - \mathbf{b}_2) \cdot \mathbf{r}) \right) -$$
$$- 4V_{1,(20)} \cos((\mathbf{b}_2 - \mathbf{b}_1) \cdot \mathbf{r}) \cos\left((\mathbf{b}_2 + \mathbf{b}_1) \cdot \mathbf{r} + \frac{2\pi}{3}\right) +$$
$$+ 2V_{1,(12)} \Big( 2\cos((\mathbf{b}_1 + 2\mathbf{b}_2) \cdot \mathbf{r}) + 2\cos((2\mathbf{b}_1 + \mathbf{b}_2) \cdot \mathbf{r}) +$$
$$+ \cos\left((\mathbf{b}_1 - 2\mathbf{b}_2) \cdot \mathbf{r} - \frac{\pi}{3}\right) - \cos\left((2\mathbf{b}_1 - \mathbf{b}_2) \cdot \mathbf{r} - \frac{2\pi}{3}\right) \Big) +$$
$$+ 2V_{1,(22)} \left( 2\cos(2(\mathbf{b}_1 - \mathbf{b}_2) \cdot \mathbf{r}) - \cos\left(2(\mathbf{b}_1 + \mathbf{b}_2) \cdot \mathbf{r} - \frac{2\pi}{3}\right) \right). \tag{19}$$

Using the vectors $\mathbf{b}_1$ and $\mathbf{b}_2$ of the reciprocal lattice from (15) and coordinates $x$ and $y$ one obtains from (19):

$$V(x, y) = 2V_{1,(00)} + 4V_{1,(10)} \cos\left(\frac{2\pi}{3a}x + \frac{\pi}{3}\right) \cos\left(\frac{2\pi}{3a}\sqrt{3}y\right) +$$
$$+ 2V_{1,(11)} \left( \cos\left(\frac{4\pi}{3a}x - \frac{\pi}{3}\right) + 2\cos\left(\frac{4\pi}{3a}\sqrt{3}y\right) \right) - 4V_{1,(20)} \cos\left(\frac{4\pi}{3a}\sqrt{3}y\right) \cos\left(\frac{4\pi}{3a}x + \frac{2\pi}{3}\right) +$$
$$+ 4V_{1,(12)} \left( 2\cos\left(\frac{2\pi}{a}x\right) \cos\left(\frac{2\pi}{3a}\sqrt{3}y\right) - \sin\left(\frac{2\pi}{3a}x - \frac{\pi}{6}\right) \cos\left(\frac{2\pi}{a}\sqrt{3}y\right) \right) +$$
$$+ 2V_{1,(22)} \left( 2\cos\left(\frac{8\pi}{3a}\sqrt{3}y\right) - \cos\left(\frac{8\pi}{3a}x - \frac{2\pi}{3}\right) \right). \tag{20}$$

We need the derivatives for the forces components in dimensionless form

$$-\frac{\partial \tilde{U}}{\partial \tilde{\xi}} = \tilde{U}'_{10} \sin\left(\frac{2\pi}{3\tilde{a}}\tilde{x} + \frac{\pi}{3}\right) \cos\left(\frac{2\pi}{3\tilde{a}}\sqrt{3}\tilde{y}\right) + \tilde{U}'_{11} \sin\left(\frac{4\pi}{3\tilde{a}}\tilde{x} - \frac{\pi}{3}\right) -$$
$$- \tilde{U}'_{20} \cos\left(\frac{4\pi}{3\tilde{a}}\sqrt{3}y\right) \sin\left(\frac{4\pi}{3\tilde{a}}x + \frac{2\pi}{3}\right) + \tilde{U}'_{12} \Bigg( 6\sin\left(\frac{2\pi}{\tilde{a}}\tilde{x}\right) \cos\left(\frac{2\pi}{3\tilde{a}}\sqrt{3}\tilde{y}\right) +$$
$$+ \cos\left(\frac{2\pi}{3\tilde{a}}\tilde{x} - \frac{\pi}{6}\right) \cos\left(\frac{2\pi}{\tilde{a}}\sqrt{3}\tilde{y}\right) \Bigg) - \tilde{U}'_{22} \sin\left(\frac{8\pi}{3\tilde{a}}\tilde{x} - \frac{2\pi}{3}\right), \tag{21}$$

$$-\frac{\partial \tilde{U}}{\partial \tilde{y}} = \tilde{U}'_{10}\sqrt{3}\cos\left(\frac{2\pi}{3\tilde{a}}\tilde{x}+\frac{\pi}{3}\right)\sin\left(\frac{2\pi}{3\tilde{a}}\sqrt{3}\tilde{y}\right) + \tilde{U}'_{11}2\sqrt{3}\sin\left(\frac{4\pi}{3\tilde{a}}\sqrt{3}\tilde{y}\right) -$$

$$-\sqrt{3}\tilde{U}'_{20}\sin\left(\frac{4\pi}{3\tilde{a}}\sqrt{3}\tilde{y}\right)\cos\left(\frac{4\pi}{3\tilde{a}}\tilde{x}+\frac{2\pi}{3}\right) + \tilde{U}'_{12}\left(2\sqrt{3}\cos\left(\frac{2\pi}{\tilde{a}}\tilde{x}\right)\sin\left(\frac{2\pi}{3\tilde{a}}\sqrt{3}\tilde{y}\right) - \right.$$

$$\left. -3\sqrt{3}\sin\left(\frac{2\pi}{3\tilde{a}}\tilde{x}-\frac{\pi}{6}\right)\sin\left(\frac{2\pi}{\tilde{a}}\sqrt{3}\tilde{y}\right)\right) + 2\sqrt{3}\tilde{U}'_{22}\sin\left(\frac{8\pi}{3\tilde{a}}\sqrt{3}\tilde{y}\right), \tag{22}$$

where the notations are introduced:

$$\tilde{U}'_{10} = \frac{8\pi}{3\tilde{a}}\tilde{V}_{1,(10)}, \quad \tilde{U}'_{11} = \frac{8\pi}{3\tilde{a}}\tilde{V}_{1,(11)}, \quad \tilde{U}'_{20} = \frac{16\pi}{3\tilde{a}}\tilde{V}_{1,(20)}, \quad \tilde{U}'_{12} = \frac{8\pi}{3\tilde{a}}\tilde{V}_{1,(12)}, \quad \tilde{U}'_{22} = \frac{16\pi}{3\tilde{a}}\tilde{V}_{1,(22)}. \tag{23}$$

Consider as the approximation the acting forces by $\tilde{t} = 0$, when $\tilde{\xi} = \tilde{x}$. After substitution of (21) and (22) in (9), one obtains the expressions for the dimensionless forces acting on the unit of mass of particles:

$$\tilde{F}_{p\xi} = -\frac{\partial \tilde{\varphi}}{\partial \tilde{\xi}} + \tilde{U}'_{10}\sin\left(\frac{2\pi}{3\tilde{a}}\tilde{\xi}+\frac{\pi}{3}\right)\cos\left(\frac{2\pi}{3\tilde{a}}\sqrt{3}\tilde{y}\right) + \tilde{U}'_{11}\sin\left(\frac{4\pi}{3\tilde{a}}\tilde{\xi}-\frac{\pi}{3}\right) -$$

$$-\tilde{U}'_{20}\cos\left(\frac{4\pi}{3\tilde{a}}\sqrt{3}\tilde{y}\right)\sin\left(\frac{4\pi}{3\tilde{a}}\tilde{\xi}+\frac{2\pi}{3}\right) + \tilde{U}'_{12}\left(6\sin\left(\frac{2\pi}{\tilde{a}}\tilde{\xi}\right)\cos\left(\frac{2\pi}{3\tilde{a}}\sqrt{3}\tilde{y}\right) + \right. \tag{24}$$

$$\left. +\cos\left(\frac{2\pi}{3\tilde{a}}\tilde{\xi}-\frac{\pi}{6}\right)\cos\left(\frac{2\pi}{\tilde{a}}\sqrt{3}\tilde{y}\right)\right) - \tilde{U}'_{22}\sin\left(\frac{8\pi}{3\tilde{a}}\tilde{\xi}-\frac{2\pi}{3}\right) + \tilde{E}_{\xi},$$

$$\tilde{F}_{py} = -\frac{\partial \tilde{\varphi}}{\partial \tilde{y}} + \tilde{U}'_{10}\sqrt{3}\cos\left(\frac{2\pi}{3\tilde{a}}\tilde{\xi}+\frac{\pi}{3}\right)\sin\left(\frac{2\pi}{3\tilde{a}}\sqrt{3}\tilde{y}\right) + \tilde{U}'_{11}2\sqrt{3}\sin\left(\frac{4\pi}{3\tilde{a}}\sqrt{3}\tilde{y}\right) -$$

$$-\sqrt{3}\tilde{U}'_{20}\sin\left(\frac{4\pi}{3\tilde{a}}\sqrt{3}\tilde{y}\right)\cos\left(\frac{4\pi}{3\tilde{a}}\tilde{\xi}+\frac{2\pi}{3}\right) + \tilde{U}'_{12}\left(2\sqrt{3}\cos\left(\frac{2\pi}{\tilde{a}}\tilde{\xi}\right)\sin\left(\frac{2\pi}{3\tilde{a}}\sqrt{3}\tilde{y}\right) - \right. \tag{25}$$

$$\left. -3\sqrt{3}\sin\left(\frac{2\pi}{3\tilde{a}}\tilde{\xi}-\frac{\pi}{6}\right)\sin\left(\frac{2\pi}{\tilde{a}}\sqrt{3}\tilde{y}\right)\right) + 2\sqrt{3}\tilde{U}'_{22}\sin\left(\frac{8\pi}{3\tilde{a}}\sqrt{3}\tilde{y}\right) + \tilde{E}_{y}.$$

Analogically

$$\tilde{F}_{e\xi} = -\tilde{F}_{p\xi}, \quad \tilde{F}_{ey} = -\tilde{F}_{py}. \tag{26}$$

The forces (24)-(26) should be introduced in the system of the hydrodynamic equations (3)-(8).

Suppose that the external field intensity **E** is equal to zero. Average on $\tilde{y}$ the obtained system of quantum hydrodynamic equations taking into account that effective hydrodynamic velocity is directed along $x$ axis. The averaging will be realized in the limit of one hexagonal crystal cell. Carry out the integration of the left and right hand sides of the hydrodynamic equations calculating the integral $\frac{1}{\sqrt{3}\tilde{a}}\int_{-\frac{\sqrt{3}}{2}\tilde{a}}^{\frac{\sqrt{3}}{2}\tilde{a}} d\tilde{y}$ (see Fig. 1) and taking into account that $\frac{1}{\sqrt{3}\tilde{a}}\int_{-\frac{\sqrt{3}}{2}\tilde{a}}^{\frac{\sqrt{3}}{2}\tilde{a}} \frac{\partial \psi}{\partial \tilde{y}} d\tilde{y} = 0$ because of system symmetry for arbitrary function $\Psi$, characterizing the state of the physical system. We suppose therefore that by averaging all physical values, characterizing the state of the physical system do not depend on $\tilde{y}$.

As result we have the following system of equations:

Dimensionless Poisson equation for the self-consistent potential $\tilde{\varphi}$ of the electric field:

$$\frac{\partial^2 \tilde{\varphi}}{\partial \tilde{\xi}^2} = -4\pi R \left\{ \frac{m_e}{m_p} \left[ \tilde{\rho}_p - \frac{m_e H}{m_p \tilde{u}^2} \frac{\partial}{\partial \tilde{\xi}} (\tilde{\rho}_p (\tilde{u}-1)) \right] - \left[ \tilde{\rho}_e - \frac{H}{\tilde{u}^2} \frac{\partial}{\partial \tilde{\xi}} (\tilde{\rho}_e (\tilde{u}-1)) \right] \right\}. \qquad (27)$$

Continuity equation for the positive particles:

$$\frac{\partial}{\partial \tilde{\xi}} [\tilde{\rho}_p (1-\tilde{u})] + \frac{m_e}{m_p} \frac{\partial}{\partial \tilde{\xi}} \left\{ \frac{H}{\tilde{u}^2} \frac{\partial}{\partial \tilde{\xi}} [\tilde{\rho}_p (\tilde{u}-1)^2] \right\} + \frac{m_e}{m_p} \frac{\partial}{\partial \tilde{\xi}} \left\{ \frac{H}{\tilde{u}^2} \left[ \frac{V_{0p}^2}{u_0^2} \frac{\partial}{\partial \tilde{\xi}} \tilde{p}_p - \right. \right.$$
$$\left. \left. -\frac{m_e}{m_p} \tilde{\rho}_p E \left( -\frac{\partial \tilde{\varphi}}{\partial \tilde{\xi}} + \tilde{U}'_{11} \sin\left(\frac{4\pi}{3\tilde{a}} \tilde{\xi} - \frac{\pi}{3}\right) - \tilde{U}'_{22} \sin\left(\frac{8\pi}{3\tilde{a}} \tilde{\xi} - \frac{2\pi}{3}\right) \right) \right] \right\} = 0 \qquad (28)$$

Continuity equation for electrons:

$$\frac{\partial}{\partial \tilde{\xi}} [\tilde{\rho}_e (1-\tilde{u})] + \frac{\partial}{\partial \tilde{\xi}} \left\{ \frac{H}{\tilde{u}^2} \frac{\partial}{\partial \tilde{\xi}} [\tilde{\rho}_e (\tilde{u}-1)^2] \right\} + \frac{\partial}{\partial \tilde{\xi}} \left\{ \frac{H}{\tilde{u}^2} \left[ \frac{V_{0e}^2}{u_0^2} \frac{\partial}{\partial \tilde{\xi}} \tilde{p}_e - \right. \right.$$
$$\left. \left. -\tilde{\rho}_e E \left( \frac{\partial \tilde{\varphi}}{\partial \tilde{\xi}} - \tilde{U}'_{11} \sin\left(\frac{4\pi}{3\tilde{a}} \tilde{\xi} - \frac{\pi}{3}\right) + \tilde{U}'_{22} \sin\left(\frac{8\pi}{3\tilde{a}} \tilde{\xi} - \frac{2\pi}{3}\right) \right) \right] \right\} = 0 \qquad (29)$$

Momentum equation for the movement along the *x* direction:

$$\frac{\partial}{\partial \tilde{\xi}} \left\{ (\tilde{\rho}_p + \tilde{\rho}_e) \tilde{u} (\tilde{u}-1) + \frac{V_{0p}^2}{u_0^2} \tilde{p}_p + \frac{V_{0e}^2}{u_0^2} \tilde{p}_e \right\} -$$
$$-\frac{m_e}{m_p} \tilde{\rho}_p E \left( -\frac{\partial \tilde{\varphi}}{\partial \tilde{\xi}} + \tilde{U}'_{11} \sin\left(\frac{4\pi}{3\tilde{a}} \tilde{\xi} - \frac{\pi}{3}\right) - \tilde{U}'_{22} \sin\left(\frac{8\pi}{3\tilde{a}} \tilde{\xi} - \frac{2\pi}{3}\right) \right) -$$
$$-\tilde{\rho}_e E \left( \frac{\partial \tilde{\varphi}}{\partial \tilde{\xi}} - \tilde{U}'_{11} \sin\left(\frac{4\pi}{3\tilde{a}} \tilde{\xi} - \frac{\pi}{3}\right) + \tilde{U}'_{22} \sin\left(\frac{8\pi}{3\tilde{a}} \tilde{\xi} - \frac{2\pi}{3}\right) \right) +$$

$$+\frac{m_e}{m_p}\frac{\partial}{\partial\tilde{\xi}}\left\{\frac{H}{\tilde{u}^2}\left[\frac{\partial}{\partial\tilde{\xi}}\left(2\frac{V_{0p}^2}{u_0^2}\tilde{p}_p(1-\tilde{u})-\tilde{\rho}_p\tilde{u}(1-\tilde{u})^2\right)-\right.\right.$$

$$\left.\left.-\frac{m_e}{m_p}\tilde{\rho}_p(1-\tilde{u})E\left(-\frac{\partial\tilde{\varphi}}{\partial\tilde{\xi}}+\tilde{U}'_{11}\sin\left(\frac{4\pi}{3\tilde{a}}\tilde{\xi}-\frac{\pi}{3}\right)-\tilde{U}'_{22}\sin\left(\frac{8\pi}{3\tilde{a}}\tilde{\xi}-\frac{2\pi}{3}\right)\right)\right]\right\}+$$

$$+\frac{\partial}{\partial\tilde{\xi}}\left\{\frac{H}{\tilde{u}^2}\left[\frac{\partial}{\partial\tilde{\xi}}\left(2\frac{V_{0e}^2}{u_0^2}\tilde{p}_e(1-\tilde{u})-\tilde{\rho}_e\tilde{u}(1-\tilde{u})^2\right)-\right.\right.$$

$$\left.\left.-\tilde{\rho}_e(1-\tilde{u})E\left(\frac{\partial\tilde{\varphi}}{\partial\tilde{\xi}}-\tilde{U}'_{11}\sin\left(\frac{4\pi}{3\tilde{a}}\tilde{\xi}-\frac{\pi}{3}\right)+\tilde{U}'_{22}\sin\left(\frac{8\pi}{3\tilde{a}}\tilde{\xi}-\frac{2\pi}{3}\right)\right)\right]\right\}+$$

$$+\frac{H}{\tilde{u}^2}E\left(\frac{m_e}{m_p}\right)^2\left(-\frac{\partial\tilde{\varphi}}{\partial\tilde{\xi}}+\tilde{U}'_{11}\sin\left(\frac{4\pi}{3\tilde{a}}\tilde{\xi}-\frac{\pi}{3}\right)-\tilde{U}'_{22}\sin\left(\frac{8\pi}{3\tilde{a}}\tilde{\xi}-\frac{2\pi}{3}\right)\right)\left(\frac{\partial}{\partial\tilde{\xi}}(\tilde{\rho}_p(\tilde{u}-1))\right)+$$

$$+\frac{H}{\tilde{u}^2}E\left(\frac{\partial\tilde{\varphi}}{\partial\tilde{\xi}}-\tilde{U}'_{11}\sin\left(\frac{4\pi}{3\tilde{a}}\tilde{\xi}-\frac{\pi}{3}\right)+\tilde{U}'_{22}\sin\left(\frac{8\pi}{3\tilde{a}}\tilde{\xi}-\frac{2\pi}{3}\right)\right)\left(\frac{\partial}{\partial\tilde{\xi}}(\tilde{\rho}_e(\tilde{u}-1))\right)-$$

$$-\frac{m_e}{m_p}\frac{\partial}{\partial\tilde{\xi}}\left\{\frac{H}{\tilde{u}^2}\frac{V_{0p}^2}{u_0^2}\frac{\partial}{\partial\tilde{\xi}}(\tilde{p}_p\tilde{u})\right\}-\frac{\partial}{\partial\tilde{\xi}}\left\{\frac{H}{\tilde{u}^2}\frac{V_{0e}^2}{u_0^2}\frac{\partial}{\partial\tilde{\xi}}(\tilde{p}_e\tilde{u})\right\}+$$

$$+\left(\frac{m_e}{m_p}\right)^2 E\frac{\partial}{\partial\tilde{\xi}}\left\{\frac{H}{\tilde{u}^2}\left[\left(-\frac{\partial\tilde{\varphi}}{\partial\tilde{\xi}}+\tilde{U}'_{11}\sin\left(\frac{4\pi}{3\tilde{a}}\tilde{\xi}-\frac{\pi}{3}\right)-\tilde{U}'_{22}\sin\left(\frac{8\pi}{3\tilde{a}}\tilde{\xi}-\frac{2\pi}{3}\right)\right)\tilde{\rho}_p\tilde{u}\right]\right\}+$$

$$+E\frac{\partial}{\partial\tilde{\xi}}\left\{\frac{H}{\tilde{u}^2}\left[\left(\frac{\partial\tilde{\varphi}}{\partial\tilde{\xi}}-\tilde{U}'_{11}\sin\left(\frac{4\pi}{3\tilde{a}}\tilde{\xi}-\frac{\pi}{3}\right)+\tilde{U}'_{22}\sin\left(\frac{8\pi}{3\tilde{a}}\tilde{\xi}-\frac{2\pi}{3}\right)\right)\tilde{\rho}_e\tilde{u}\right]\right\}=0 \quad (30)$$

Energy equation for the positive particles:

$$\frac{\partial}{\partial\tilde{\xi}}\left[\tilde{\rho}_p\tilde{u}^2(\tilde{u}-1)+5\frac{V_{0p}^2}{u_0^2}\tilde{p}_p\tilde{u}-3\frac{V_{0p}^2}{u_0^2}\tilde{p}_p\right]-$$

$$-2\frac{m_e}{m_p}\tilde{\rho}_p E\left(-\frac{\partial\tilde{\varphi}}{\partial\tilde{\xi}}+\tilde{U}'_{11}\sin\left(\frac{4\pi}{3\tilde{a}}\tilde{\xi}-\frac{\pi}{3}\right)-\tilde{U}'_{22}\sin\left(\frac{8\pi}{3\tilde{a}}\tilde{\xi}-\frac{2\pi}{3}\right)\right)\tilde{u}+$$

$$+\frac{\partial}{\partial\tilde{\xi}}\left\{\frac{H}{\tilde{u}^2}\frac{m_e}{m_p}\left[\frac{\partial}{\partial\tilde{\xi}}\left(-\tilde{\rho}_p\tilde{u}^2(1-\tilde{u})^2+7\frac{V_{0p}^2}{u_0^2}\tilde{p}_p\tilde{u}(1-\tilde{u})+3\frac{V_{0p}^2}{u_0^2}\tilde{p}_p(\tilde{u}-1)-\frac{V_{0p}^2}{u_0^2}\tilde{p}_p\tilde{u}^2-5\frac{V_{0p}^4}{u_0^4}\frac{\tilde{p}_p^2}{\tilde{\rho}_p}\right)+\right.\right.$$

$$\left.\left.+E\left(-2\frac{m_e}{m_p}\tilde{\rho}_p\tilde{u}(1-\tilde{u})+\frac{m_e}{m_p}\tilde{\rho}_p\tilde{u}^2+5\frac{m_e}{m_p}\frac{V_{0p}^2}{u_0^2}\tilde{p}_p\right)\left(-\frac{\partial\tilde{\varphi}}{\partial\tilde{\xi}}+\right.\right.\right.$$

$$\left.\left.\left.+\tilde{U}'_{11}\sin\left(\frac{4\pi}{3\tilde{a}}\tilde{\xi}-\frac{\pi}{3}\right)-\tilde{U}'_{22}\sin\left(\frac{8\pi}{3\tilde{a}}\tilde{\xi}-\frac{2\pi}{3}\right)\right)\right]\right\}+2\frac{H}{\tilde{u}^2}E\left(\frac{m_e}{m_p}\right)^2\left[-\frac{\partial}{\partial\tilde{\xi}}(\tilde{\rho}_p\tilde{u}(1-\tilde{u}))+\right.$$

$$\left.+\frac{V_{0p}^2}{u_0^2}\frac{\partial}{\partial\tilde{\xi}}\tilde{p}_p\right]\left(-\frac{\partial\tilde{\varphi}}{\partial\tilde{\xi}}+\tilde{U}'_{11}\sin\left(\frac{4\pi}{3\tilde{a}}\tilde{\xi}-\frac{\pi}{3}\right)-\tilde{U}'_{22}\sin\left(\frac{8\pi}{3\tilde{a}}\tilde{\xi}-\frac{2\pi}{3}\right)\right)-$$

$$-2\frac{H}{\tilde{u}^2}E^2\left(\frac{m_e}{m_p}\right)^3\tilde{\rho}_p\left[\left(-\frac{\partial\tilde{\varphi}}{\partial\tilde{\xi}}+\tilde{U}'_{11}\sin\left(\frac{4\pi}{3\tilde{a}}\tilde{\xi}-\frac{\pi}{3}\right)-\tilde{U}'_{22}\sin\left(\frac{8\pi}{3\tilde{a}}\tilde{\xi}-\frac{2\pi}{3}\right)\right)^2+\right.$$

$$+\frac{1}{2}\left(\tilde{U}'_{10}\sin\left(\frac{2\pi}{3\tilde{a}}\tilde{\xi}+\frac{\pi}{3}\right)+6\tilde{U}'_{12}\sin\left(\frac{2\pi}{\tilde{a}}\tilde{\xi}\right)\right)^2+\frac{1}{2}(\tilde{U}'_{12})^2\cos^2\left(\frac{2\pi}{3\tilde{a}}\tilde{\xi}-\frac{\pi}{6}\right)+$$

$$+\frac{1}{2}(\tilde{U}'_{02})^2\sin^2\left(\frac{4\pi}{3\tilde{a}}\tilde{\xi}+\frac{2\pi}{3}\right)-\frac{4}{3\pi}\tilde{U}'_{02}\sin\left(\frac{4\pi}{3\tilde{a}}\tilde{\xi}+\frac{2\pi}{3}\right)\left(\tilde{U}'_{10}\sin\left(\frac{2\pi}{3\tilde{a}}\tilde{\xi}+\frac{\pi}{3}\right)+6\tilde{U}'_{12}\sin\left(\frac{2\pi}{\tilde{a}}\tilde{\xi}\right)\right)-$$

$$-\frac{12}{5\pi}\tilde{U}'_{02}\tilde{U}'_{12}\sin\left(\frac{4\pi}{3\tilde{a}}\tilde{\xi}+\frac{2\pi}{3}\right)\cos\left(\frac{2\pi}{3\tilde{a}}\tilde{\xi}-\frac{\pi}{6}\right)+\frac{3}{2}\left(\tilde{U}'_{10}\cos\left(\frac{2\pi}{3\tilde{a}}\tilde{\xi}+\frac{\pi}{3}\right)+2\tilde{U}'_{12}\cos\left(\frac{2\pi}{\tilde{a}}\tilde{\xi}\right)\right)^2+$$

$$+\frac{3}{2}\left(2\tilde{U}'_{11}-\tilde{U}'_{02}\cos\left(\frac{4\pi}{3\tilde{a}}\tilde{\xi}+\frac{2\pi}{3}\right)\right)^2+\frac{27}{2}(\tilde{U}'_{12})^2\sin^2\left(\frac{2\pi}{3\tilde{a}}\tilde{\xi}-\frac{\pi}{6}\right)+6(\tilde{U}'_{22})^2+$$

$$+\frac{8}{\pi}\left(\tilde{U}'_{10}\cos\left(\frac{2\pi}{3\tilde{a}}\tilde{\xi}+\frac{\pi}{3}\right)+2\tilde{U}'_{12}\cos\left(\frac{2\pi}{\tilde{a}}\tilde{\xi}\right)\right)\left(2\tilde{U}'_{11}-\tilde{U}'_{02}\cos\left(\frac{4\pi}{3\tilde{a}}\tilde{\xi}+\frac{2\pi}{3}\right)\right)-$$

$$-\frac{96}{15\pi}\left(\tilde{U}'_{10}\cos\left(\frac{2\pi}{3\tilde{a}}\tilde{\xi}+\frac{\pi}{3}\right)+2\tilde{U}'_{12}\cos\left(\frac{2\pi}{\tilde{a}}\tilde{\xi}\right)\right)\tilde{U}'_{22}-$$

$$-\frac{72}{5\pi}\tilde{U}'_{12}\left(2\tilde{U}'_{11}-\tilde{U}'_{02}\cos\left(\frac{4\pi}{3\tilde{a}}\tilde{\xi}+\frac{2\pi}{3}\right)\right)\sin\left(\frac{2\pi}{3\tilde{a}}\tilde{\xi}-\frac{\pi}{6}\right)-\frac{288}{7\pi}\tilde{U}'_{12}\tilde{U}'_{22}\sin\left(\frac{2\pi}{3\tilde{a}}\tilde{\xi}-\frac{\pi}{6}\right)\right]=$$

$$=-\frac{\tilde{u}^2}{Hu_0^2}\left(V_{0p}^2\tilde{\rho}_p-\tilde{p}_e V_{0e}^2\right)\left(1+\frac{m_p}{m_e}\right) \qquad (31)$$

Energy equation for electrons:

$$\frac{\partial}{\partial\tilde{\xi}}\left[\tilde{\rho}_e\tilde{u}^2(\tilde{u}-1)+5\frac{V_{0e}^2}{u_0^2}\tilde{p}_e\tilde{u}-3\frac{V_{0e}^2}{u_0^2}\tilde{p}_e\right]-$$

$$-2\tilde{\rho}_e\tilde{u}E\left(\frac{\partial\tilde{\varphi}}{\partial\tilde{\xi}}-\tilde{U}'_{11}\sin\left(\frac{4\pi}{3\tilde{a}}\tilde{\xi}-\frac{\pi}{3}\right)+\tilde{U}'_{22}\sin\left(\frac{8\pi}{3\tilde{a}}\tilde{\xi}-\frac{2\pi}{3}\right)\right)+$$

$$+\frac{\partial}{\partial\tilde{\xi}}\left\{\frac{H}{\tilde{u}^2}\left[\frac{\partial}{\partial\tilde{\xi}}\left(-\tilde{\rho}_e\tilde{u}^2(1-\tilde{u})^2+7\frac{V_{0e}^2}{u_0^2}\tilde{p}_e\tilde{u}(1-\tilde{u})+3\frac{V_{0e}^2}{u_0^2}\tilde{p}_e(\tilde{u}-1)-\frac{V_{0e}^2}{u_0^2}\tilde{p}_e\tilde{u}^2-5\frac{V_{0e}^4}{u_0^4}\frac{\tilde{p}_e^2}{\tilde{\rho}_e}\right)+\right.\right.$$

$$\left.\left.+E\left(-2\tilde{\rho}_e\tilde{u}(1-\tilde{u})+\tilde{\rho}_e\tilde{u}^2+5\frac{V_{0e}^2}{u_0^2}\tilde{p}_e\right)\left(\frac{\partial\tilde{\varphi}}{\partial\tilde{\xi}}-\tilde{U}'_{11}\sin\left(\frac{4\pi}{3\tilde{a}}\tilde{\xi}-\frac{\pi}{3}\right)+\tilde{U}'_{22}\sin\left(\frac{8\pi}{3\tilde{a}}\tilde{\xi}-\frac{2\pi}{3}\right)\right)\right]\right\}+$$

$$+E\left(-2\frac{H}{\tilde{u}^2}\frac{\partial}{\partial\tilde{\xi}}(\tilde{\rho}_e\tilde{u}(1-\tilde{u}))+2\frac{H}{\tilde{u}^2}\frac{V_{0e}^2}{u_0^2}\frac{\partial}{\partial\tilde{\xi}}\tilde{p}_e\right)\left(\frac{\partial\tilde{\varphi}}{\partial\tilde{\xi}}-\tilde{U}'_{11}\sin\left(\frac{4\pi}{3\tilde{a}}\tilde{\xi}-\frac{\pi}{3}\right)+\tilde{U}'_{22}\sin\left(\frac{8\pi}{3\tilde{a}}\tilde{\xi}-\frac{2\pi}{3}\right)\right)-$$

$$-2E^2 \frac{H}{\tilde{u}^2}\tilde{\rho}_e \left[ \left( -\frac{\partial \tilde{\varphi}}{\partial \tilde{\xi}} + \tilde{U}'_{11}\sin\left(\frac{4\pi}{3\tilde{a}}\tilde{\xi} - \frac{\pi}{3}\right) - \tilde{U}'_{22}\sin\left(\frac{8\pi}{3\tilde{a}}\tilde{\xi} - \frac{2\pi}{3}\right) \right)^2 + \right.$$

$$+\frac{1}{2}\left( \tilde{U}'_{10}\sin\left(\frac{2\pi}{3\tilde{a}}\tilde{\xi} + \frac{\pi}{3}\right) + 6\tilde{U}'_{12}\sin\left(\frac{2\pi}{\tilde{a}}\tilde{\xi}\right) \right)^2 + \frac{1}{2}\left(\tilde{U}'_{12}\right)^2 \cos^2\left(\frac{2\pi}{3\tilde{a}}\tilde{\xi} - \frac{\pi}{6}\right) +$$

$$+\frac{1}{2}\left(\tilde{U}'_{02}\right)^2 \sin^2\left(\frac{4\pi}{3\tilde{a}}\tilde{\xi} + \frac{2\pi}{3}\right) - \frac{4}{3\pi}\tilde{U}'_{02}\sin\left(\frac{4\pi}{3\tilde{a}}\tilde{\xi} + \frac{2\pi}{3}\right)\left(\tilde{U}'_{10}\sin\left(\frac{2\pi}{3\tilde{a}}\tilde{\xi} + \frac{\pi}{3}\right) + 6\tilde{U}'_{12}\sin\left(\frac{2\pi}{\tilde{a}}\tilde{\xi}\right)\right) -$$

$$-\frac{12}{5\pi}\tilde{U}'_{02}\tilde{U}'_{12}\sin\left(\frac{4\pi}{3\tilde{a}}\tilde{\xi} + \frac{2\pi}{3}\right)\cos\left(\frac{2\pi}{3\tilde{a}}\tilde{\xi} - \frac{\pi}{6}\right) + \frac{3}{2}\left(\tilde{U}'_{10}\cos\left(\frac{2\pi}{3\tilde{a}}\tilde{\xi} + \frac{\pi}{3}\right) + 2\tilde{U}'_{12}\cos\left(\frac{2\pi}{\tilde{a}}\tilde{\xi}\right)\right)^2 +$$

$$+\frac{3}{2}\left(2\tilde{U}'_{11} - \tilde{U}'_{02}\cos\left(\frac{4\pi}{3\tilde{a}}\tilde{\xi} + \frac{2\pi}{3}\right)\right)^2 + \frac{27}{2}\left(\tilde{U}'_{12}\right)^2 \sin^2\left(\frac{2\pi}{3\tilde{a}}\tilde{\xi} - \frac{\pi}{6}\right) + 6\left(\tilde{U}'_{22}\right)^2 +$$

$$+\frac{8}{\pi}\left(\tilde{U}'_{10}\cos\left(\frac{2\pi}{3\tilde{a}}\tilde{\xi} + \frac{\pi}{3}\right) + 2\tilde{U}'_{12}\cos\left(\frac{2\pi}{\tilde{a}}\tilde{\xi}\right)\right)\left(2\tilde{U}'_{11} - \tilde{U}'_{02}\cos\left(\frac{4\pi}{3\tilde{a}}\tilde{\xi} + \frac{2\pi}{3}\right)\right) -$$

$$-\frac{96}{15\pi}\left(\tilde{U}'_{10}\cos\left(\frac{2\pi}{3\tilde{a}}\tilde{\xi} + \frac{\pi}{3}\right) + 2\tilde{U}'_{12}\cos\left(\frac{2\pi}{\tilde{a}}\tilde{\xi}\right)\right)\tilde{U}'_{22} -$$

$$\left. -\frac{72}{5\pi}\tilde{U}'_{12}\left(2\tilde{U}'_{11} - \tilde{U}'_{02}\cos\left(\frac{4\pi}{3\tilde{a}}\tilde{\xi} + \frac{2\pi}{3}\right)\right)\sin\left(\frac{2\pi}{3\tilde{a}}\tilde{\xi} - \frac{\pi}{6}\right) - \frac{288}{7\pi}\tilde{U}'_{12}\tilde{U}'_{22}\sin\left(\frac{2\pi}{3\tilde{a}}\tilde{\xi} - \frac{\pi}{6}\right) \right] =$$

$$= -\frac{\tilde{u}^2}{Hu_0^2}\left(V_{0e}^2 \tilde{\rho}_e - V_{0p}^2 \tilde{\rho}_p\right)\left(1 + \frac{m_p}{m_e}\right) \tag{32}$$

### 3. Estimations of the numerical parameters.

We need estimations for the numerical values of dimensionless parameters for solutions of the hydrodynamic equations (27) - (32). In its turn these parameters depend on choosing of the independent scales physical values. Analyze the independent scales for the physical problem under consideration.

The surface electron density in graphene is about $\breve{n}_e \approx 10^{10}\,см^{-2}$, the thickness of the graphene layer is equal to ~ 1 $nm$. Then the electron concentration consists $n_e \approx 10^{17}\,cm^{-3}$, and the density for the electron species $\rho_e = m_e n_e \approx 10^{-10}\,g/cm^3$ which leads to the scale $\rho_0 = 10^{-10}\,g/cm^3$. For numerical solutions of the hydrodynamic equations (27)-(32) we need Cauchy conditions, obviously in the typical for graphene conditions the estimation $\tilde{\rho}_e \sim 1$ is valid which can be used as the condition by $\tilde{\xi} = 0$.

The process of the carbon atoms polarization leads to displacement of the atoms from the regular chain and to the creation of the "effective" positive particles which concentration $n_p \approx n_e$. Masses of these particles is about the mass of the carbon atom $m_p \approx 2 \cdot 10^{-23}\,г$. Then, $\frac{L}{T} = \frac{m_e}{m_p} \approx 5 \cdot 10^{-5}$; $\rho_p = m_p n_p \approx 2 \cdot 10^{-6}\,g/cm^3$ and by the choosed scale for the density $\rho_0$ we have $\tilde{\rho}_p \sim 2 \cdot 10^4$.

Going to the scales for thermal velocities for electrons and the positive particles we have by $T=300°K$:

$$V_{0e} \sim \sqrt{\frac{k_B T}{m_e}} \approx 6.4 \cdot 10^6 \ см/с, \text{ take the scale } V_{0e} = 5 \cdot 10^6 \ см/с;$$

$$V_{0p} \sim \sqrt{\frac{k_B T}{m_p}} \approx 4.5 \cdot 10^4 \ см/с, \text{ take the scale } V_{0p} = 5 \cdot 10^4 \ см/с.$$

The theoretical mobility in graphene reaches up to $10^6 \ cm^2/V \cdot s$. Let us use the scale

$$u_0 = 5 \cdot 10^6 \ cm/s. \text{ Then } N = \frac{V_{0e}^2}{u_0^2} = 1, \ P = \frac{V_{0p}^2}{u_0^2} = 10^{-4}.$$

Let us estimate the parameters $E$ and $R$. For this estimation we need the scale $\varphi_0$. Admit $\varphi_0 \approx \delta \frac{e}{a}$, where $\delta$ is a "shielding coefficient". Naturally to take $x_0 = a = 0.142 \ nm$ (see Fig. 1) - as the length scale, then $\tilde{a} = 1$. In the situation of a uncertainty in $\varphi_0$ choosing let us consider two limit cases:

1) $\delta \sim 1$.

Then $E = \frac{e\varphi_0}{m_e u_0^2} \sim 1000, \ R = \frac{e\rho_0 x_0^2}{m_e \varphi_0} \sim 3 \cdot 10^{-7}$.

2) $\delta = 0.0001$.

Then $E = \frac{e\varphi_0}{m_e u_0^2} \sim 0.1, \ R = \frac{e\rho_0 x_0^2}{m_e \varphi_0} \sim 3 \cdot 10^{-3}$.

Consider the terms describing the lattice influence. We should estimate the coefficients (23) using $\varphi_0$ as the scale for the potential $V$, $V = \varphi_0 \tilde{V}$. Three possible cases under consideration:

1) $V \sim \varphi_0$

We choose $U = \tilde{U}'_{10} \sim 10, \ F = \tilde{U}'_{11} \sim 10, \ J = \tilde{U}'_{20} \sim \pm 5, \ B = \tilde{U}'_{12} \sim \pm 2,5, \ G = \tilde{U}'_{22} \sim \pm 5$.

In this case the coefficients of "the second order" are less than the coefficients of "the first order."

2) $V \prec\prec \varphi_0$ (The small influence of the lattice),

We choose $U = \tilde{U}'_{10} \sim 0.1, \ F = \tilde{U}'_{11} \sim 0.1, \ J = \tilde{U}'_{20} \sim 0.05, \ B = \tilde{U}'_{12} \sim 0.025, \ G = \tilde{U}'_{22} \sim 0.05$.

3) $V \succ\succ \varphi_0$ (The great influence of the lattice),

We choose $U = \tilde{U}'_{10} \sim 1000, \ F = \tilde{U}'_{11} \sim 1000, \ J = \tilde{U}'_{20} \sim 500, \ B = \tilde{U}'_{12} \sim 250, \ G = \tilde{U}'_{22} \sim 500$.

Estimate parameter $H = \frac{N_R \hbar}{m_e x_0 u_0}$ for two limit cases:

1) $N_R = 1$, then $H \sim 15$.

2) $N_R = 100$, then $H \sim 1500$.

Initial conditions demand also the estimations for the quantum electron pressure and the pressure for the positive species. For the electron pressure we have $p_e = \rho_0 V_{0e}^2 \tilde{p}_e$ and using for the scale estimation $p_e = n_e k_B T \sim n_e m_e V_{oe}^2 = \rho_e V_{oe}^2 \sim \rho_0 V_{oe}^2$, one obtains $\tilde{p}_e \sim 1$. Analogically for the positive particles $p_p = \rho_0 V_{0p}^2 \tilde{p}_p$, and using $p_p = n_p k_B T \sim n_p m_p V_{op}^2 = \rho_p V_{0p}^2$, we have $p_p \sim 2 \cdot 10^4 \rho_0 V_{0p}^2$, $\tilde{p}_p \sim 2 \cdot 10^4$.

Tables 1, 2 contain the initial conditions and parameters which were not varied by the numerical modeling.

Table 1. Initial conditions.

| $\tilde{\rho}_e(0)$ | $\tilde{\rho}_p(0)$ | $\tilde{\varphi}(0)$ | $\tilde{p}_e(0)$ | $\tilde{p}_p(0)$ | $\frac{\partial \tilde{\rho}_e}{\partial \tilde{\xi}}(0)$ | $\frac{\partial \tilde{\rho}_p}{\partial \tilde{\xi}}(0)$ | $\frac{\partial \tilde{\varphi}}{\partial \tilde{\xi}}(0)$ | $\frac{\partial \tilde{p}_e}{\partial \tilde{\xi}}(0)$ | $\frac{\partial \tilde{p}_p}{\partial \tilde{\xi}}(0)$ |
|---|---|---|---|---|---|---|---|---|---|
| 1 | $2 \cdot 10^4$ | 1 | 1 | $2 \cdot 10^4$ | 0 | 0 | 0 | 0 | 0 |

Table 2. Constant parameters.

| $\tilde{a}$ | $L$ | $T$ | $N$ | $P$ |
|---|---|---|---|---|
| 1 | 1 | 20000 | 1 | $10^{-4}$ |

Table 3 contains parameters (for the six different cases) which were varied by the numerical modeling.

Table 3. Varied parameters.

| Variant № | $E$ | $R$ | $H$ | $U$ | $F$ | $J$ | $B$ | $G$ |
|---|---|---|---|---|---|---|---|---|
| 1 | 0.1 | 0.003 | 15 | 10 | 10 | 5 | 2.5 | 5 |
| 2 | 0.1 | 0.003 | 15 | 0.1 | 0.1 | 0.05 | 0.025 | 0.05 |
| 3 | 0.1 | 0.003 | 15 | 10 | 10 | -5 | -2.5 | -5 |
| 4 | 1000 | $3 \cdot 10^{-7}$ | 15 | 10 | 10 | 5 | 2.5 | 5 |
| 5 | 0.1 | 0.003 | 1500 | 10 | 10 | 5 | 2.5 | 5 |
| 6 | 0.1 | 0.003 | 15 | 1000 | 1000 | 500 | 250 | 500 |

In the present time there no the foolproof methods of the calculations of the potential lattice forces in graphene. In the following mathematical modeling the strategy is taken consisting in the vast variation of the parameters defining the evolution of the physical system.

**4. Results of the mathematical modeling without the external electric field.**

The calculations are realized on the basement of equations (27)-(32) by the initial conditions and parameters containing in the Tables 1 – 3. Now we are ready to display the results of the mathematical modeling realized with the help of Maple (the versions Maple 9 or more can be used). The system of generalized hydrodynamic equations (27) – (32) have the great possibilities of mathematical modeling as result of changing of Cauchy conditions and parameters describing the character features of initial perturbations which lead to the soliton formation.

The following Maple notations on figures are used: r- density $\tilde{\rho}_p$, s - density $\tilde{\rho}_e$, u- velocity $\tilde{u}$ ( solid black line), p - pressure $\tilde{p}_p$ (black dashed line), q – pressure $\tilde{p}_e$ and v - self consistent potential $\tilde{\varphi}$. Explanations placed under all following figures, Maple program contains Maple's notations – for example, the expression $D(u)(0) = 0$ means in the usual notations $\frac{\partial \tilde{u}}{\partial \tilde{\xi}}(0) = 0$, independent variable $t$ responds to $\tilde{\xi}$.

Important to underline that no special boundary conditions were used for all following cases. The aim of the numerical investigation consists in the discovery of the soliton waves as a product of the self-organization of matter in graphene. It means that the solution should exist only in the restricted domain of the 1D space and the obtained object in the moving coordinate system ($\tilde{\xi} = \tilde{x} - \tilde{t}$) has the constant velocity $\tilde{u} = 1$ for all parts of the object. In this case the domain of the solution existence defines the character soliton size. The following numerical results demonstrate the realization of mentioned principles.

Figures 2 - 9 reflect the result of calculations for Variant 1 (Table 3) in the first and the second approximations. In the first approximation the terms of series (18) with $|g_1| \leq 1$, $|g_2| \leq 1$ (then coefficients *U* and *F*) were taken into account. The second approximation contains all terms of the series (18) with $|g_1| \leq 2$, $|g_2| \leq 2$ (then coefficients *U*, *F*, *J*, *B* and *G*).

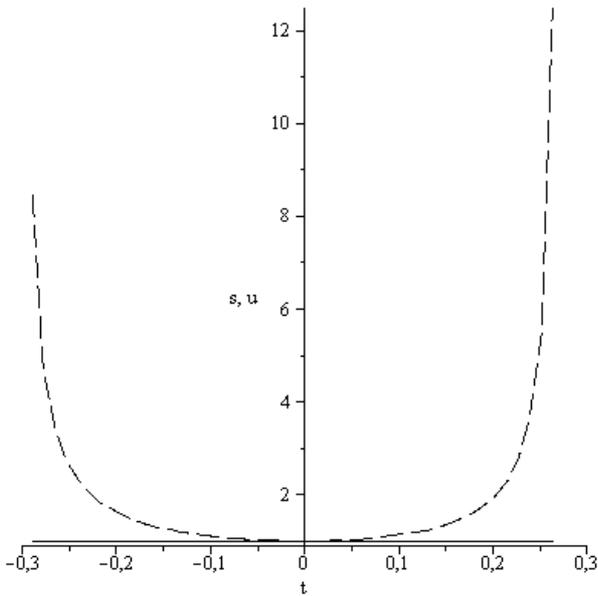

Fig. 2. s – the electron density $\tilde{\rho}_e$,
u – velocity $\tilde{u}$ (solid line).
(first approximation, Variant 1).

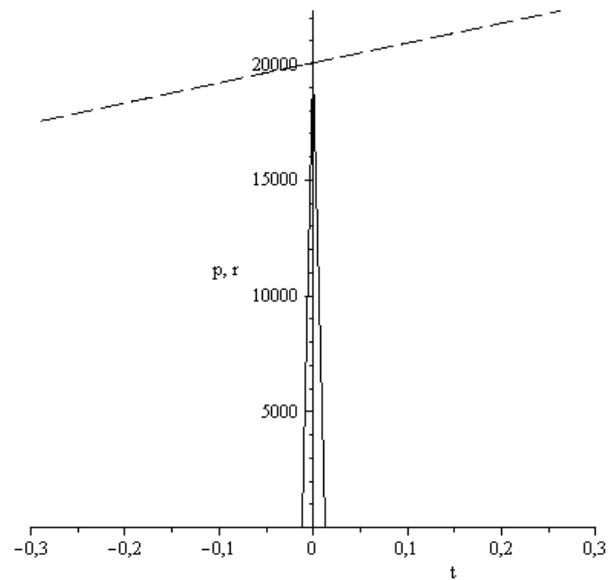

Fig. 3. r – the positive particles density,
(solid line); p – the positive particles pressure
(first approximation, Variant 1)

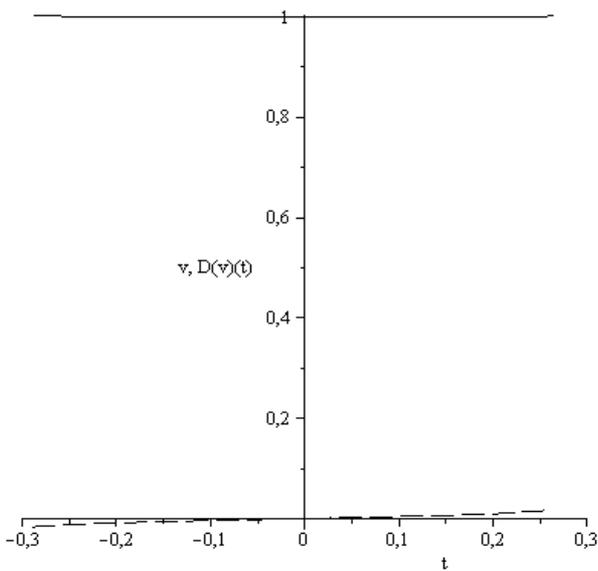

Fig. 4. v – potential $\tilde{\varphi}$ (solid line).
and derivative $D(v)(t)$.
(first approximation, Variant 1).

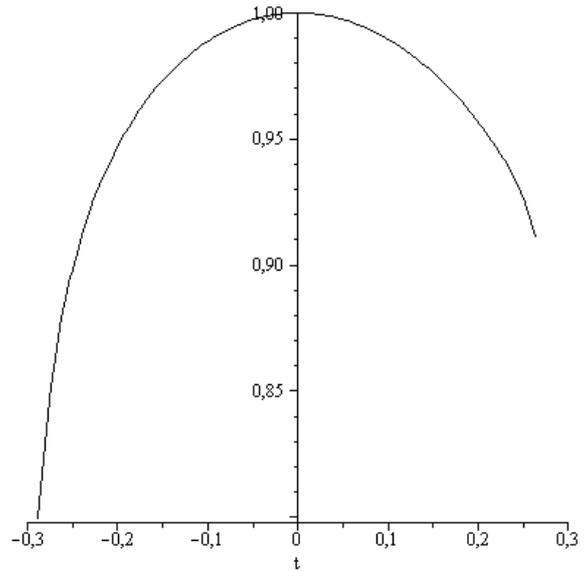

Fig. 5. q – pressure of the negative particles.
(first approximation, Variant 1).

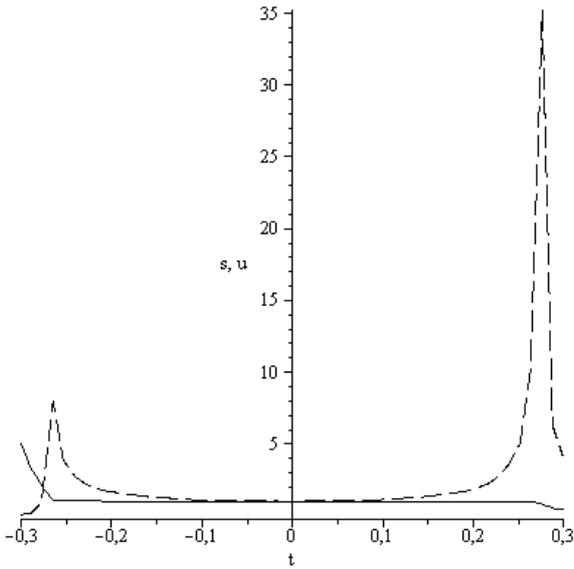

Fig. 6. s – electron density $\tilde{\rho}_e$,
u – velocity $\tilde{u}$ (solid line),
(the second approximation, Variant 1).

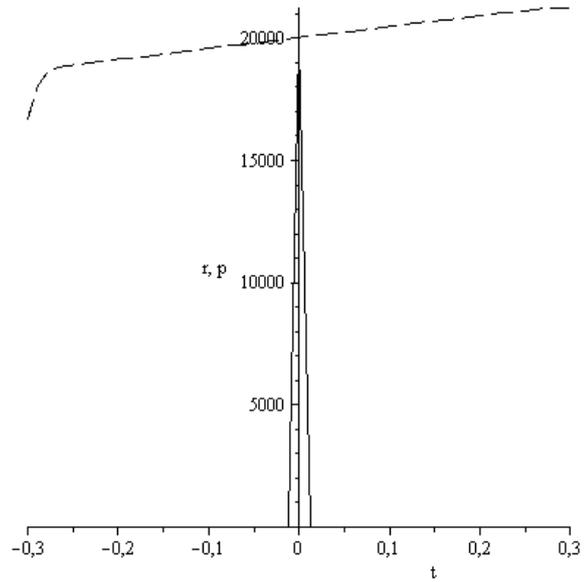

Fig. 7. r – the positive particles density (solid line)
p – the positive particles pressure,
(the second approximation, Variant 1).

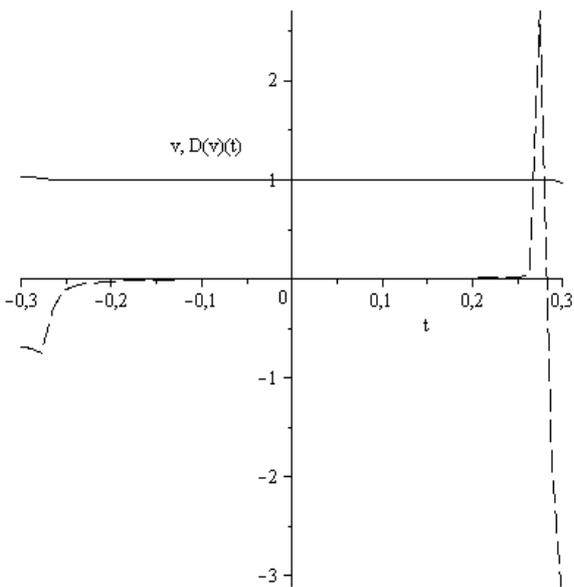

Fig. 8. v – potential $\tilde{\varphi}$ (solid line),
and derivative D(v)(t).
(the second approximation, Variant 1).

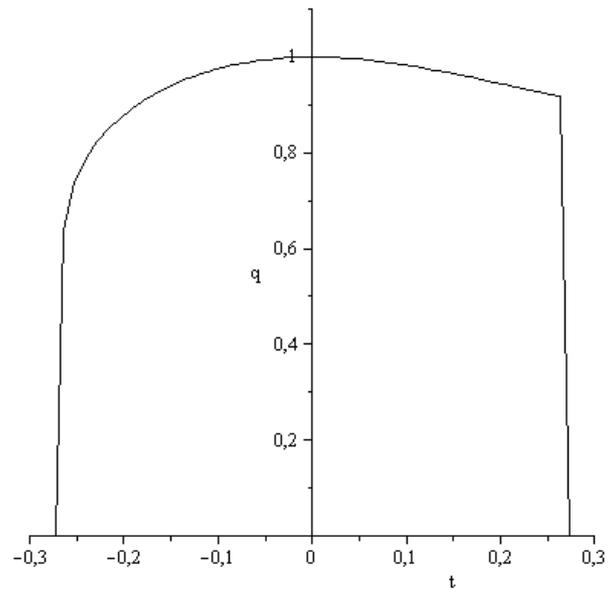

Fig. 9. q – the negative particles pressure.
(the second approximation, Variant 1).

From figures 2 - 9 follow that the size of the created soliton is about 0.5*a*, where *a*=0.142 *nm*. The domain size occupied by the polarized positive charge is about 0.025*a* (see Figs. 3, 7). But the negative charge distributes over the entire soliton domain (Figs. 2, 6), but the negative charge density increases to the edges of the soliton. Therefore the soliton structure reminds the 1D atom with the positive nuclei and the negative shell.

The self-consistent potential $\tilde{\varphi}$ is practically constant in the soliton boundaries, (Figs. 4, 8). The small grows of the positive particles pressure exists in the $x$ direction. This effect can be connected with the hydrodynamic movement along $x$ and "the reconstruction" of the polarized particles in the soliton front.

Comparing the figures 2 – 5 and 6 – 9 we conclude that the calculation results in the first and the second approximation do not vary significantly. Seemingly significant difference of figures 2 and 6 on the edges of the domain has not the physical sense because corresponds to the regions where $u \neq const$. Then the restriction of two successive approximations is justified. Along with it the question about the convergence of the series lives open because the first and the second approximations include only the restricted quantity of terms of the infinite series with the coefficients known with the small accuracy.

Figures 10 - 15 show the results of calculations responding to Variant 3 (Table 3). In the first approximation Variant 3 is identical to Variant 1 (coefficients $J = B = G = 0$) and only the results of the second approximation are delivered. These calculations are more complicated in the numerical realization and all curves are imaged separately, (Figures 10 – 15).

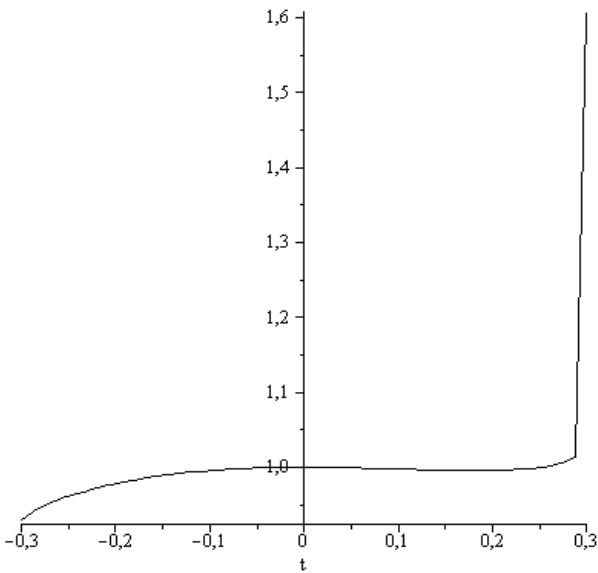 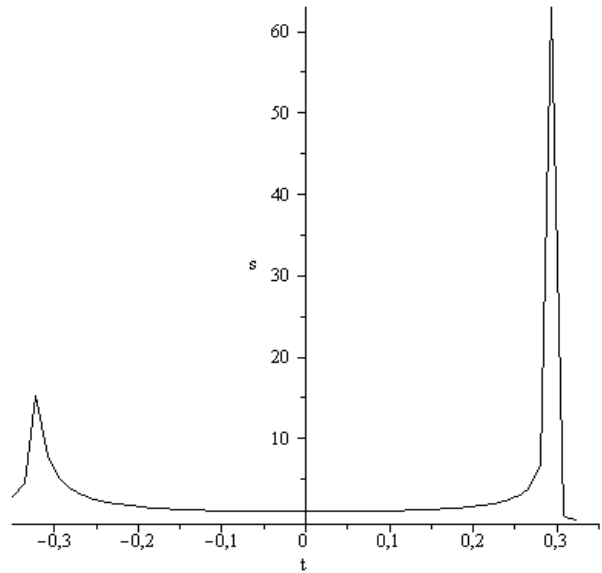

Fig. 10. u – velocity $\tilde{u}$ .                              Fig. 11. s – electron density $\tilde{\rho}_e$,
(the second approximation, Variant 3).         (the second approximation, Variant 3).

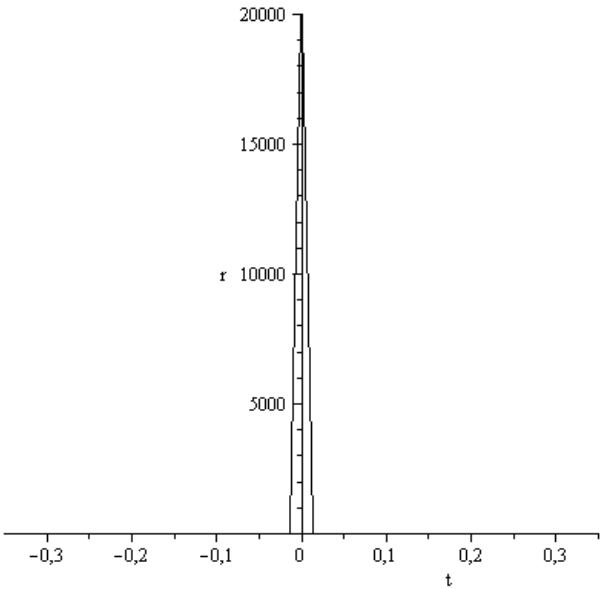

Fig. 12. r – the positive particles density.
(the second approximation, Variant 3).

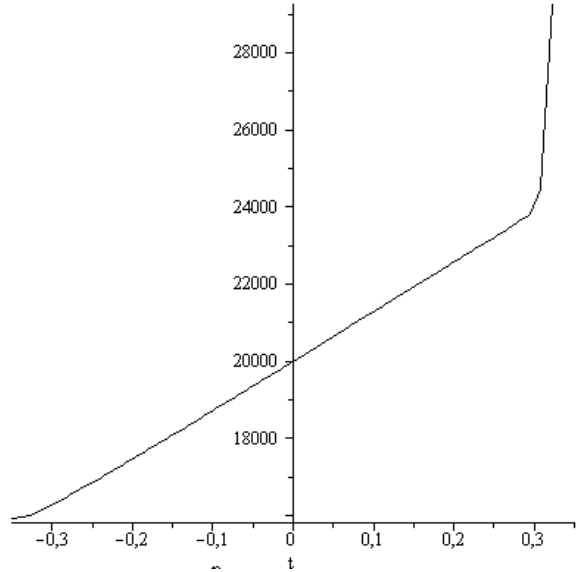

Fig. 13. p – the positive particles pressure,
(the second approximation, Variant 3).

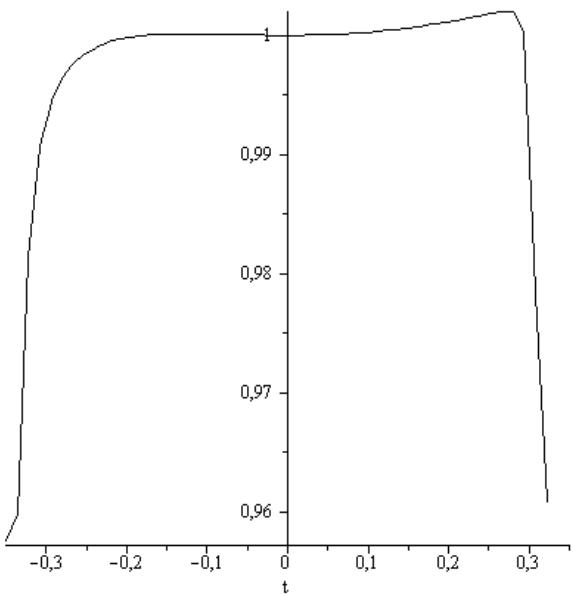

Fig. 14. v – potential $\tilde{\varphi}$.
(the second approximation, Variant 3).

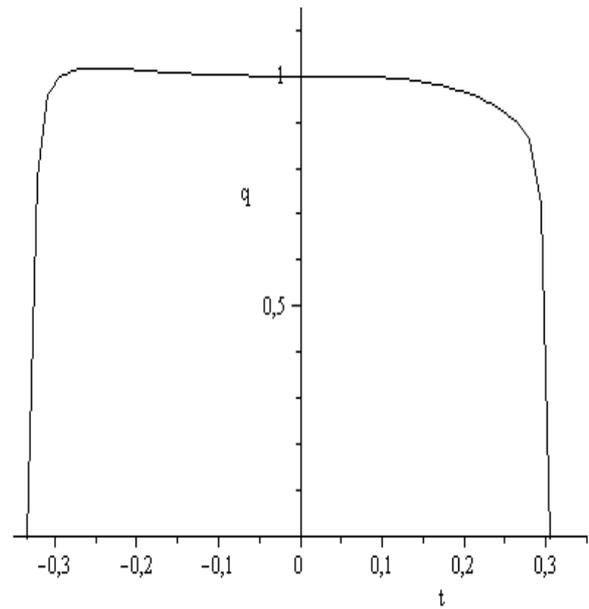

Fig. 15. q – the negative particles pressure.
(the second approximation, Variant 3).

In the comparison with Variant 1 the calculations in Variant 3 are realized for the case with opposite signs in front of the coefficients of second order. In this case the distortion of the left side of soliton is observed because by $\tilde{\tilde{\xi}} \prec 0$ the velocity $\tilde{u}$ is not constant. Then this kind of potential for lattice is not favorable for creation of the super-conducting structures.

Variant 2 (Table 3) correspond to diminishing of the lattice potential in 100 times by the same practically self-consistent potential, (see figures 16 – 23).

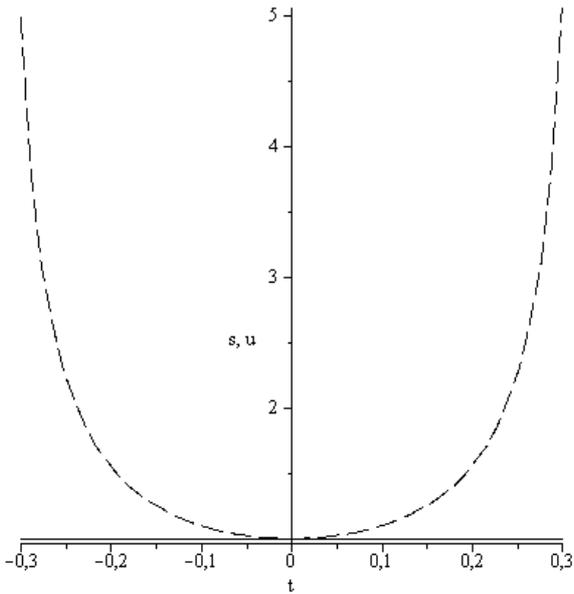

Fig. 16. s – electron density $\tilde{\rho}_e$,,
u – velocity $\tilde{u}$ (solid line).
(the first approximation, Variant 2).

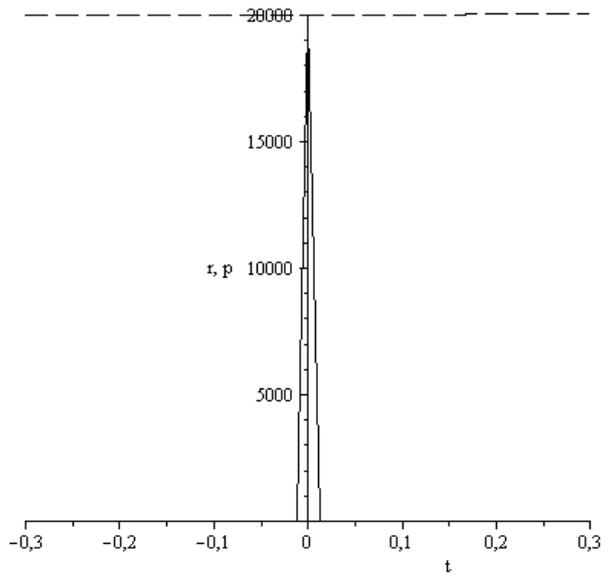

Fig. 17. r – the positive particles density,
(solid line); p – the positive particles pressure
(the first approximation, Variant 2).

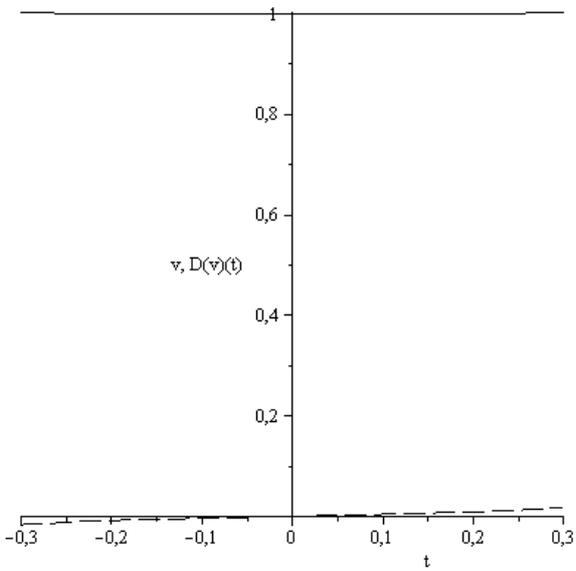

Fig. 18. v – potential $\tilde{\varphi}$ (solid line),
D(v)(t),(the first approximation, Variant 2).

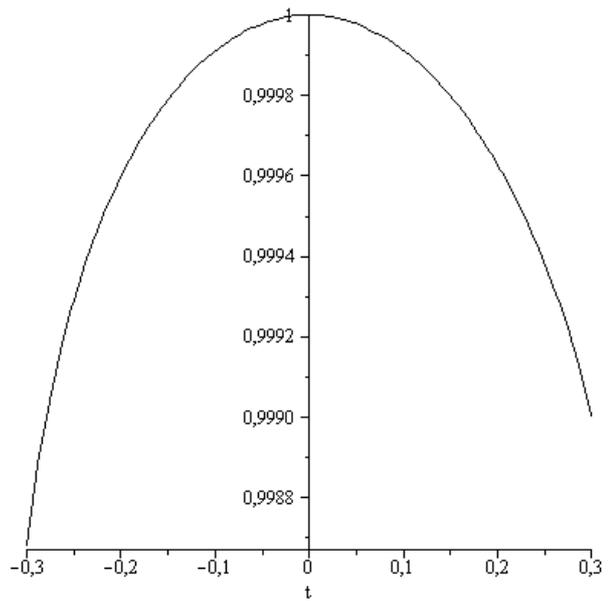

Fig. 19. q – the negative particles pressure.
(the first approximation, Variant 2).

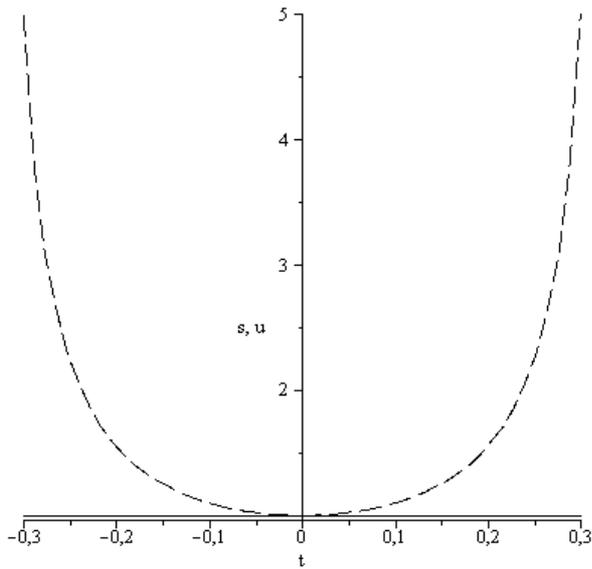

Fig. 20. s – electron density $\tilde{\rho}_e$,

u – velocity $\tilde{u}$ (solid line).

(the second approximation, Variant 2).

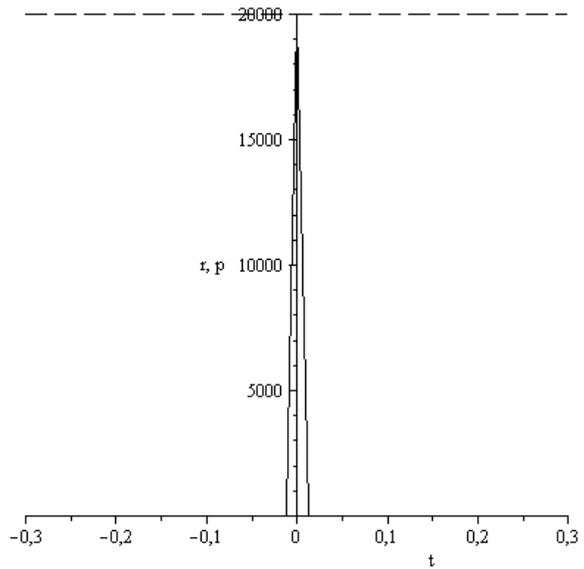

Fig. 21. r – the positive particles density,

(solid line); p – the positive particles pressure

(the second approximation, Variant 2).

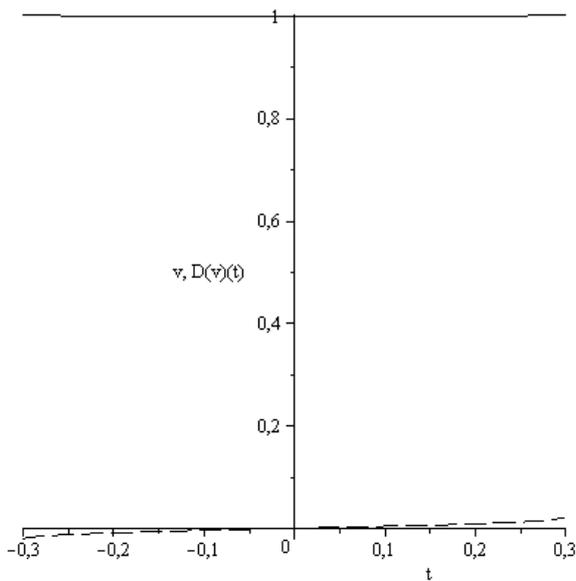

Fig. 22. v – potential $\tilde{\varphi}$ (solid line),

D(v)(t).

(the second approximation, Variant 2).

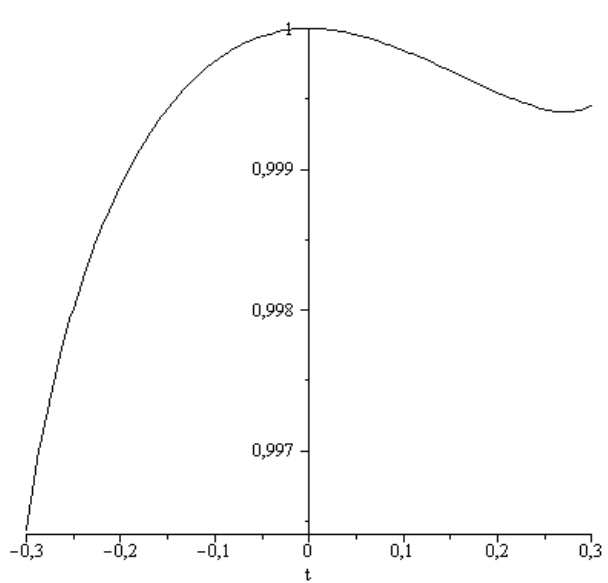

Fig. 23. q – the negative particles pressure.

(the second approximation, Variant 2).

From comparison of figures 2 - 9 and 16 - 23 follow that numerical diminishing of the lattice potential (by the practically the same value of the self-consistent potential) does not influence on soliton size. But at the same time the solitons gain the more symmetrical forms. Therefore namely the self-consistent potential plays the basic role in the soliton formation.

Let us analyze now the influence of $H$ - parameter, practically the influence of the non-locality parameter. Figures 24 – 31 (Variant 5) correspond to increasing of the parameter $H$ in 100 times in comparison with Variant 1.

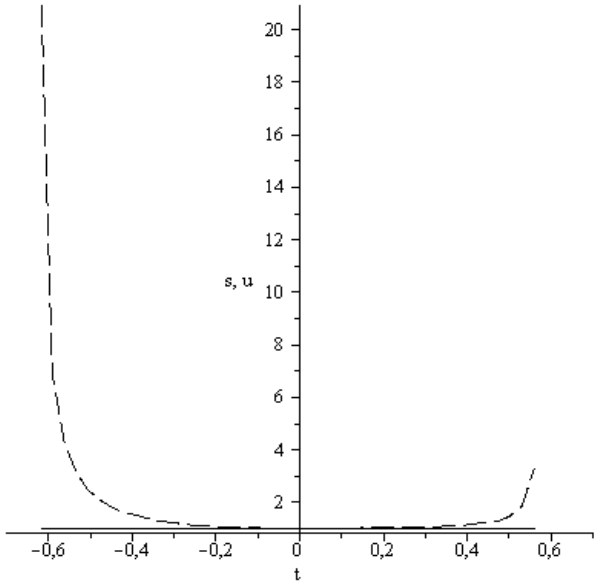

Fig. 24. s – electron density $\tilde{\rho}_e$,
u – velocity $\tilde{u}$ (solid line).
(the first approximation, Variant 5).

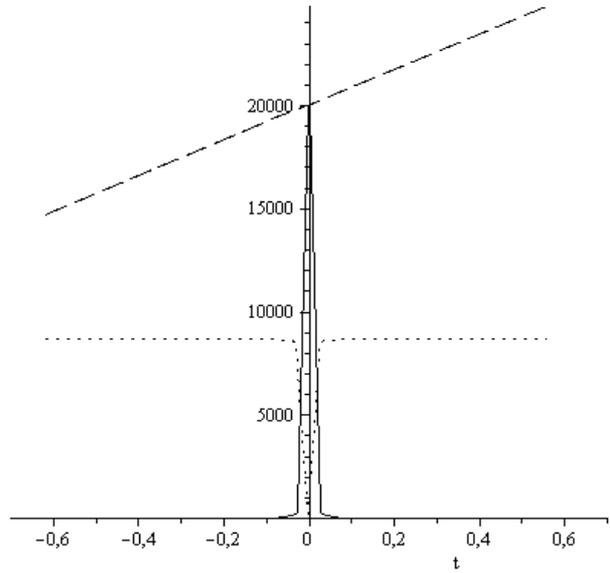

Fig. 25. r – the positive particles density, (solid line); p – the positive particles pressure (dashed line), D(p)(t) - dotted line. (the first approximation, Variant 5).

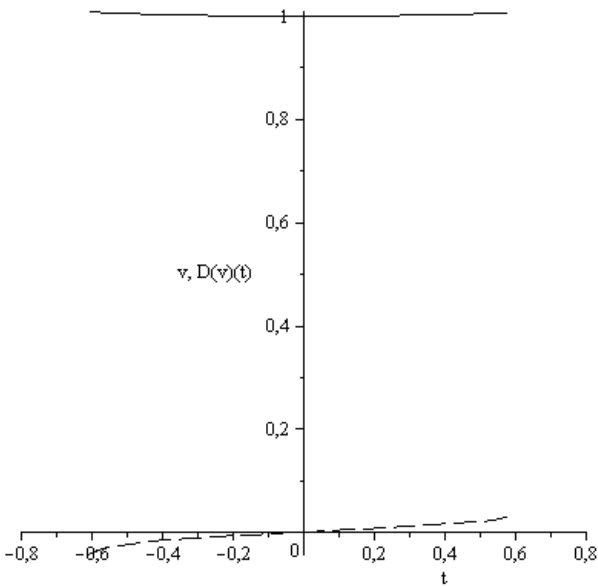

Fig. 26. v – potential $\tilde{\varphi}$ (solid line); $D(v)(t)$, (the first approximation, Variant 5).

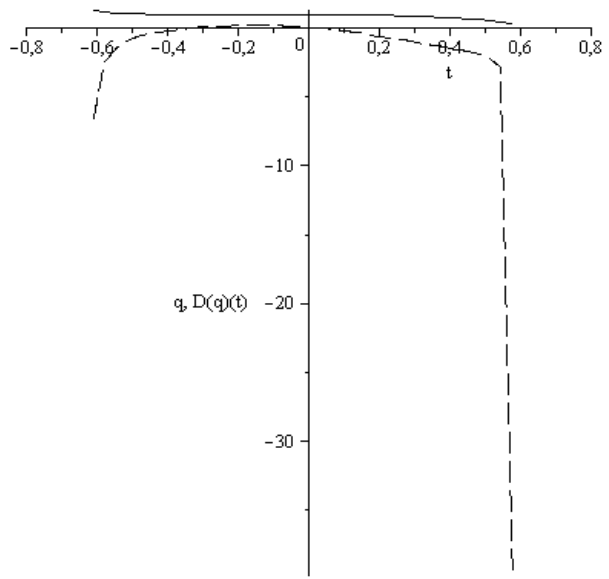

Fig. 27. q – the negative particles pressure. (solid line), $D(q)(t)$, (the first approximation, Variant 5)

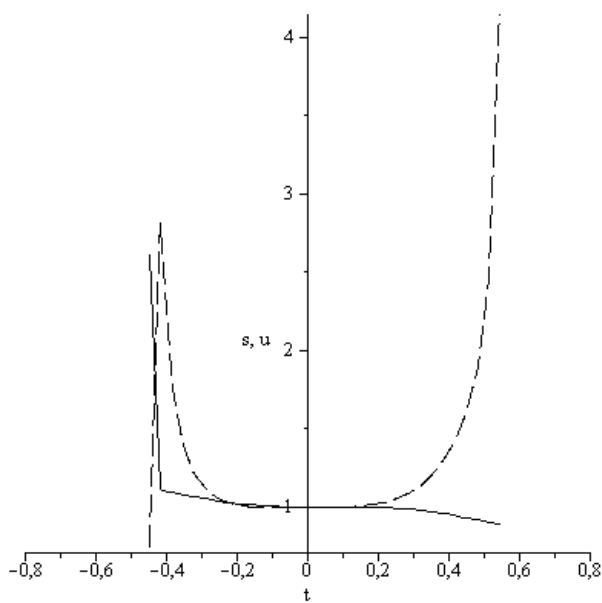

Fig. 28. s – electron density $\tilde{\rho}_e$,
u – velocity $\tilde{u}$ (solid line).
(the second approximation, Variant 5)

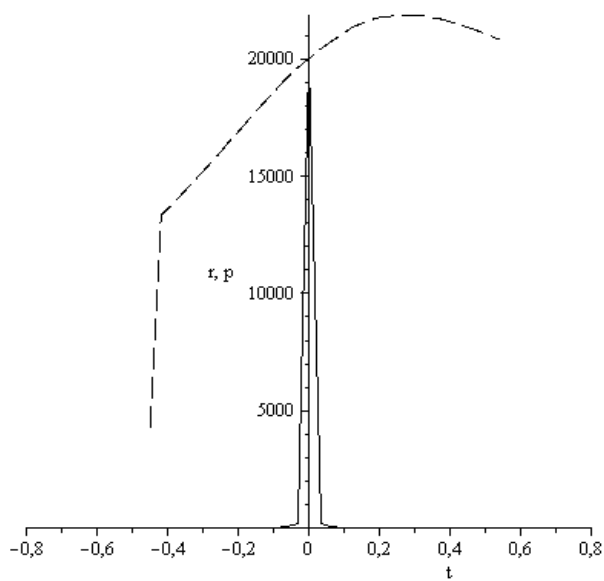

Fig. 29. r – the positive particles density,
(solid line); p – the positive particles pressure
(the second approximation, Variant 5).

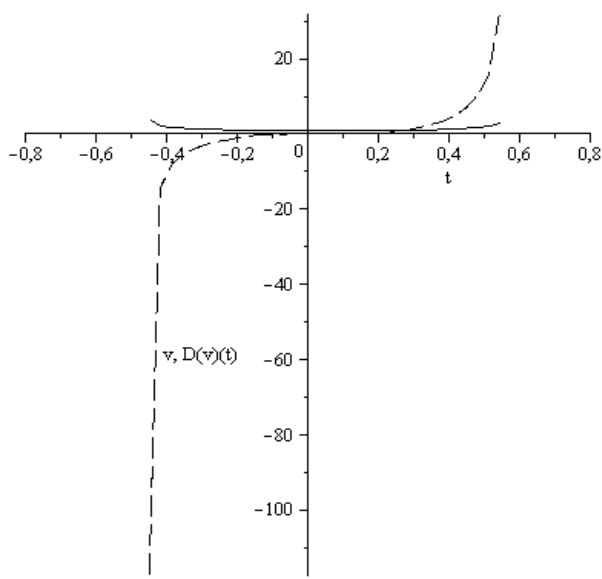

Fig. 30. v – potential $\tilde{\varphi}$ (solid line);
D(v)(t), (the second approximation, Variant 5).

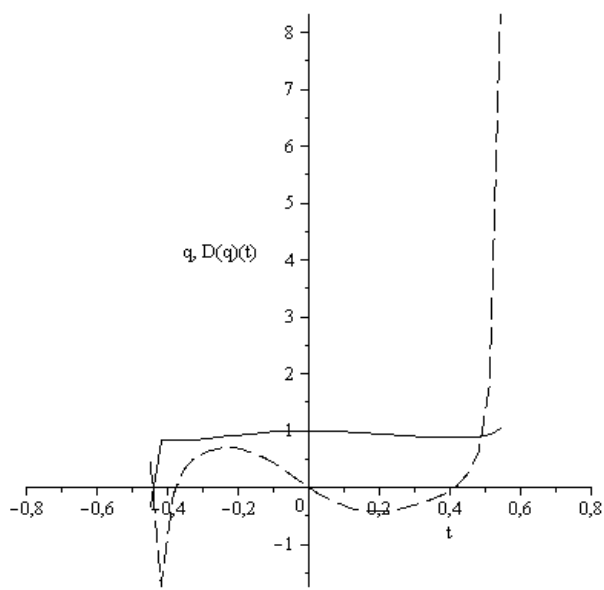

Fig. 31. q – the negative particles pressure
(solid line), D(q)(t), (the second approximation,
Variant 5).

The comparison of figures 2 - 5 and 24 - 27 indicates that in the first approximation the very significant increasing in of the *H* value in 100 times leads to increasing of the soliton size only in two times without significant changing of the soliton structure. The comparison of calculations (see figures 6 and 28) in the second approximation leads to conclusion that the region (where the velocity $\tilde{u}$ is constant) has practically the same size.

Consider now the calculations responding to Variant 4 (Table 3). Increasing in $10^4$ times of the scale $\varphi_0$ denotes increasing the self consistent potential and the lattice potential introduced in the process of the mathematical modeling. This case leads to the drastic diminishing of the soliton size. Figures 32 - 35 demonstrate that in the calculations of the first approximation the soliton size is $\sim 10^{-4} a = 1.42 \cdot 10^{-12} cm$ and exceeds the nuclei size only in several times. The positive kernel of the soliton decreasing in the less degree and occupies now the half of the soliton size. It is no surprise because the low boundary of this kernel size is the character size of the nuclei. Application of the second approximation for the lattice potential function in the mathematical modeling leads to the significant soliton deformation but the same soliton size (see figures 36-39).

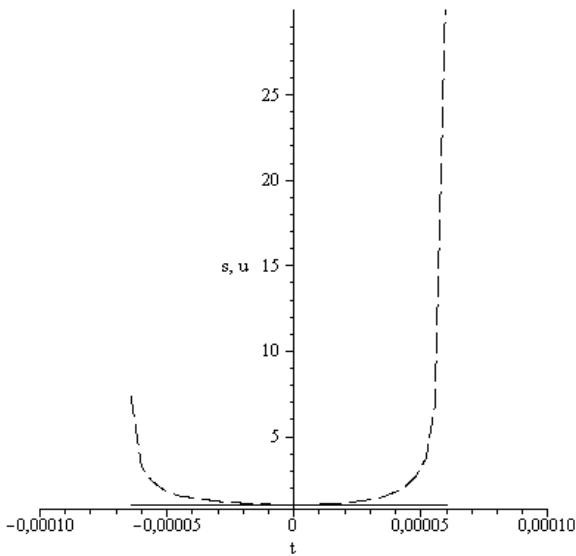

Fig. 32. s – electron density $\tilde{\rho}_e$,
u – velocity $\tilde{u}$ (solid line).
(the first approximation, Variant 4).

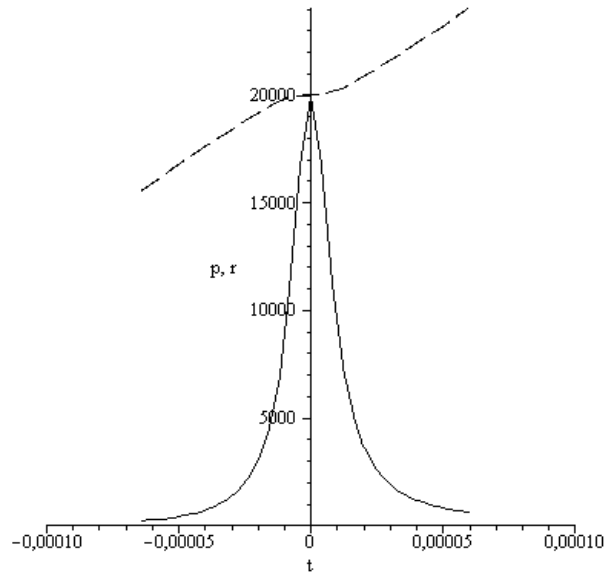

Fig. 33. r – the positive particles density,
(solid line); p – the positive particles pressure
(the first approximation, Variant 4).

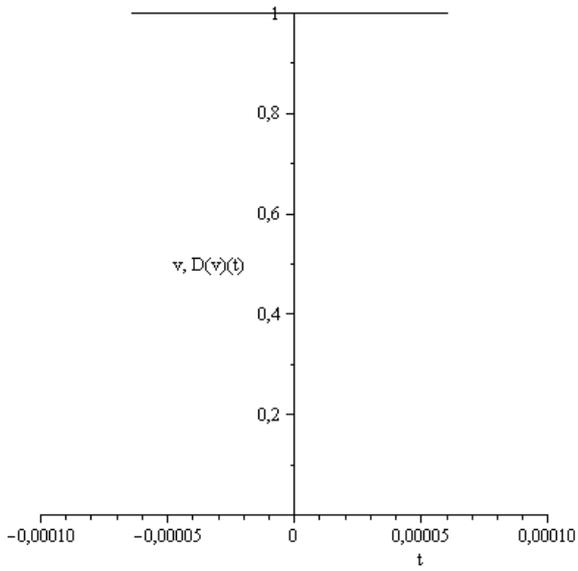

Fig. 34. v – potential $\tilde{\varphi}$ (solid line).
(the first approximation, Variant 4).

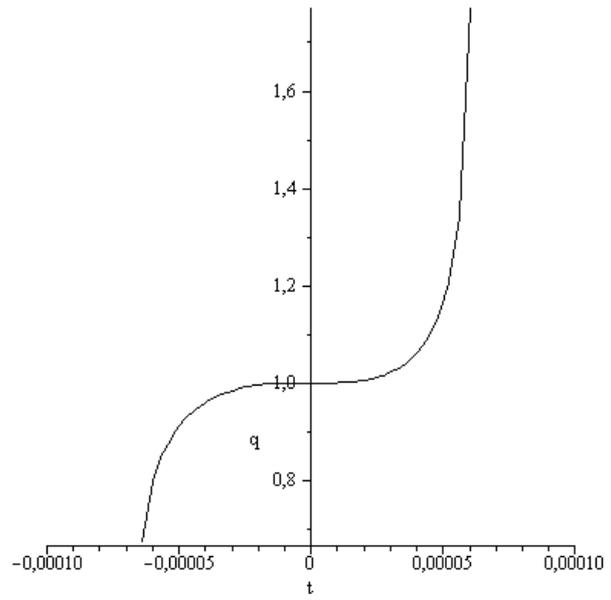

Fig. 35. q – the negative particles pressure.
(the first approximation, Variant 4).

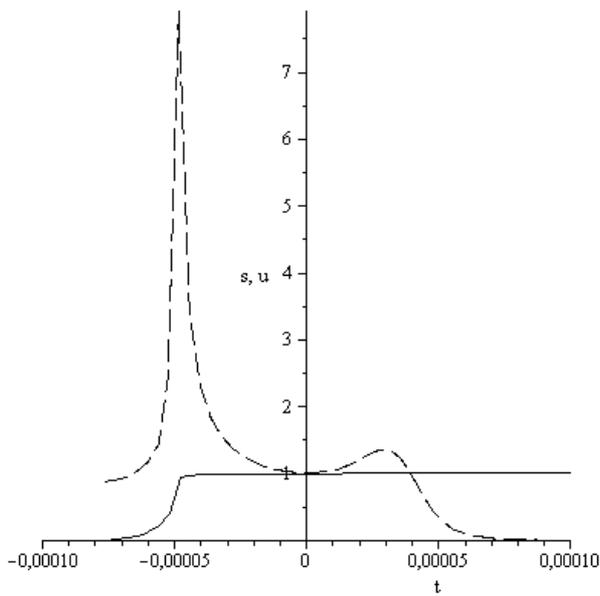

Fig. 36. s – electron density $\tilde{\rho}_e$,
u – velocity $\tilde{u}$ (solid line).
(the second approximation, Variant 4).

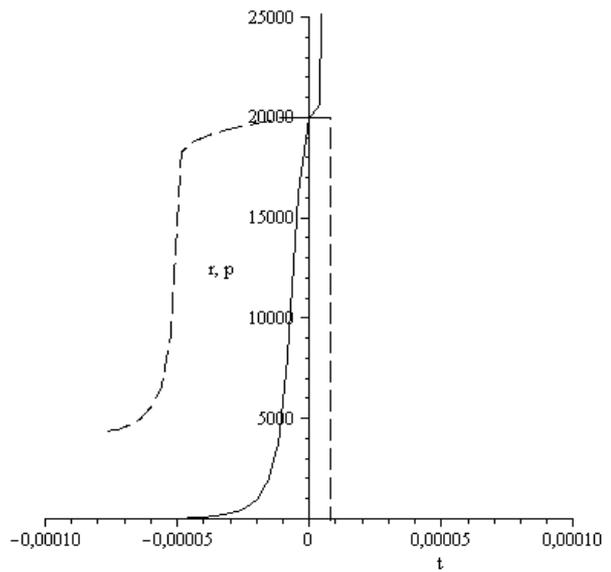

Fig. 37. r – the positive particles density,
(solid line); p – the positive particles pressure
(the second approximation, Variant 4)

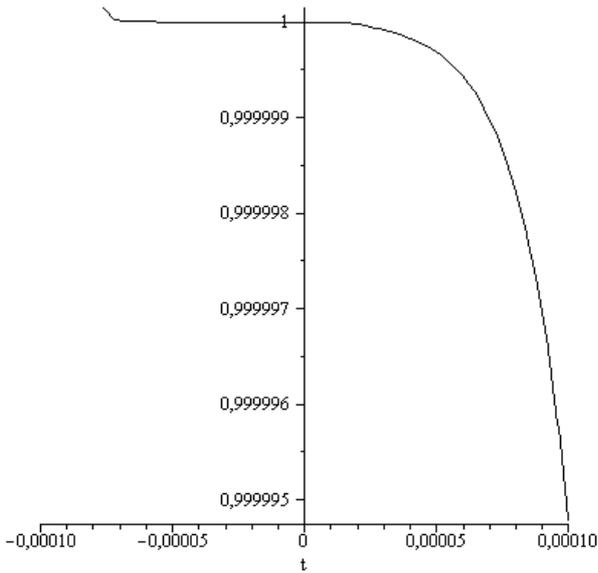

Fig. 38. v – potential $\tilde{\varphi}$ (solid line).
(the second approximation, Variant 4)

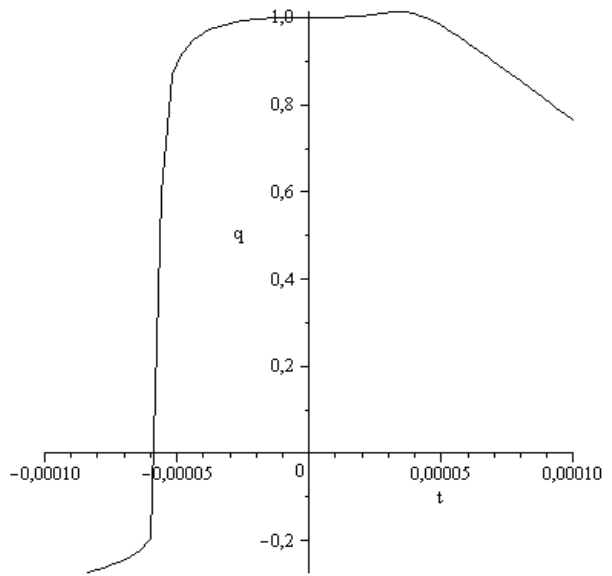

Fig. 39. q – the negative particles pressure.
(the second approximation, Variant 4)

The drastic increasing of the periodic potential of the crystal lattice (in hundred times, see figures 40 – 48) in comparison with the self-consistent potential also leads to diminishing of the soliton size. For the case Variant 6, Table 3 this size consists only $\sim 10^{-2} a$. But this increasing does not lead to the relative increasing of the soliton kernel and to the mentioned above the soliton deformation in the second approximation (see figures 45 – 48). Figure 41 demonstrate the extremely high accuracy of the soliton stability, the velocity fluctuation inside the soliton is only $\sim 10^{-16} \tilde{u}$.

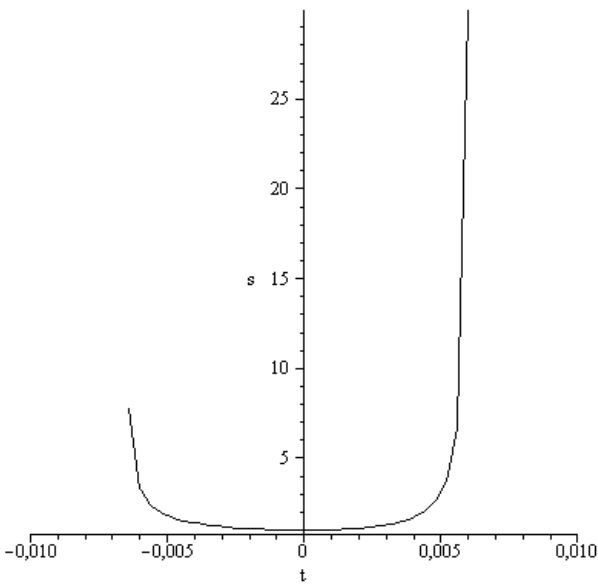

Fig. 40. s – electron density $\tilde{\rho}_e$,
(the first approximation, Variant 6).

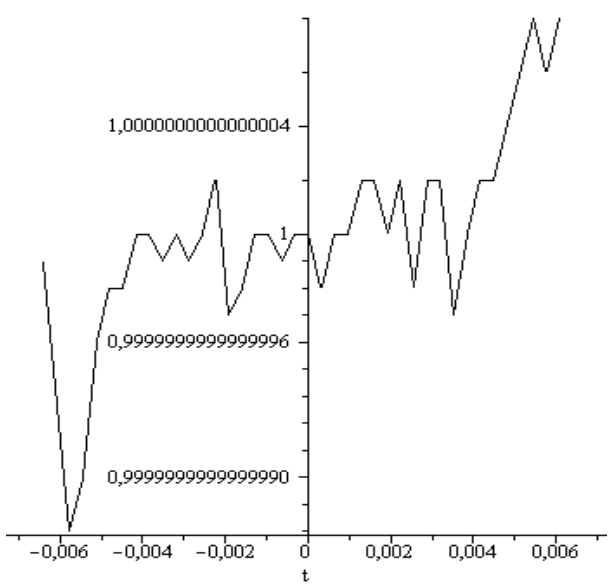

Fig. 41. u – velocity $\tilde{u}$.
(the first approximation, Variant 6).

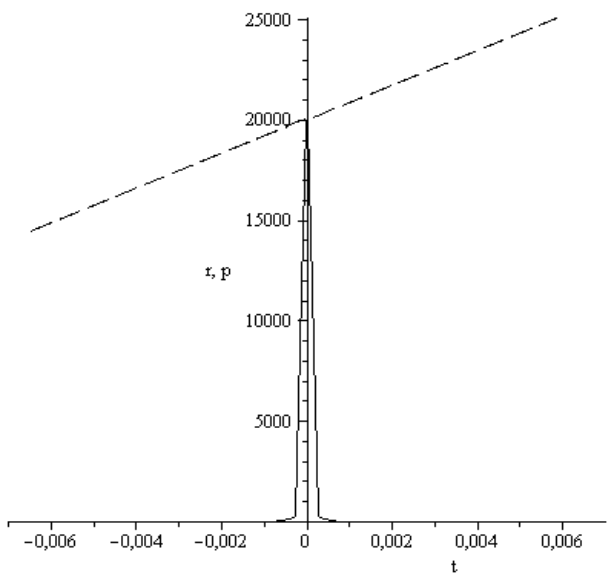

Fig. 42. r – the positive particles density, (solid line); p – the positive particles pressure (the first approximation, Variant 6).

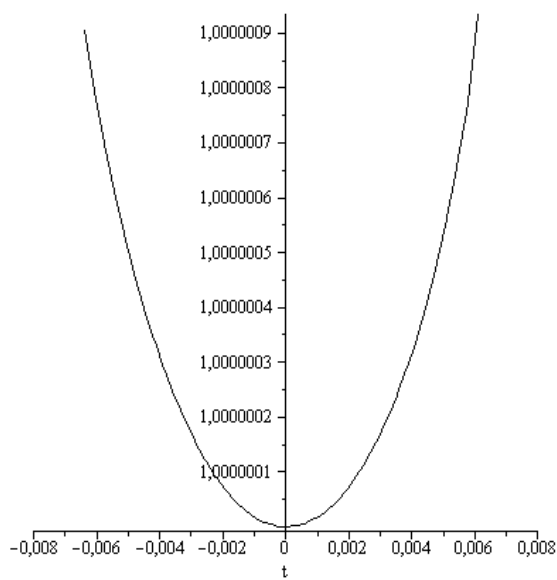

Fig. 43. v – potential $\tilde{\varphi}$. (the first approximation, Variant 6).

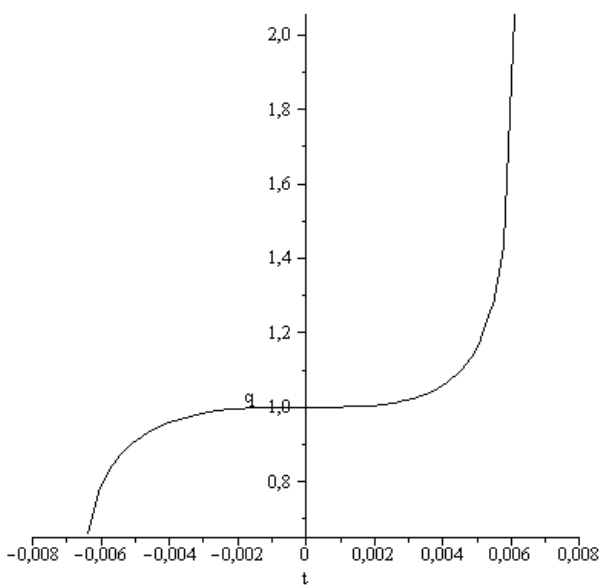

Fig. 44. q – the negative particles pressure. (the first approximation, Variant 6)..

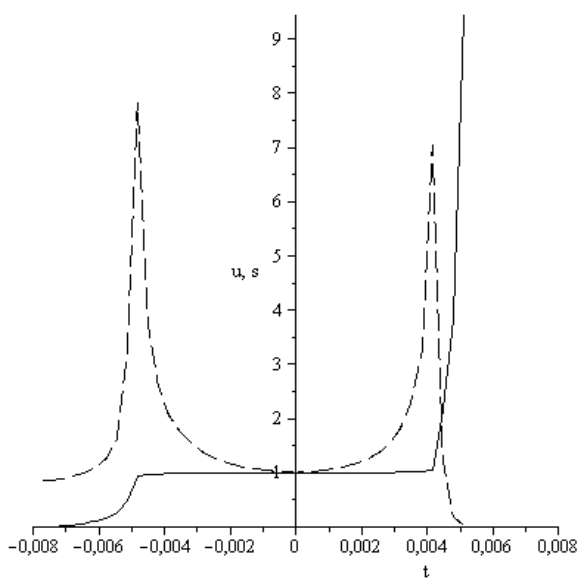

Fig. 45. s – electron density $\tilde{\rho}_e$, u – velocity $\tilde{u}$ (solid line). (the second approximation, Variant 6).

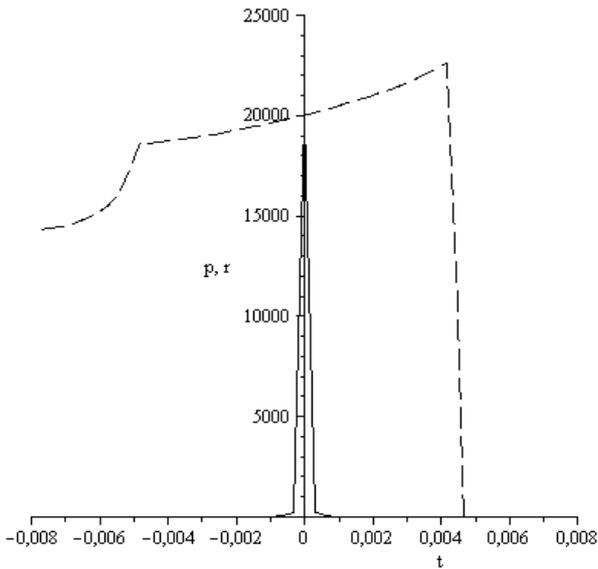

Fig. 46. r – the positive particles density. (solid line); p – the positive particles pressure, (the second approximation, Variant 6).

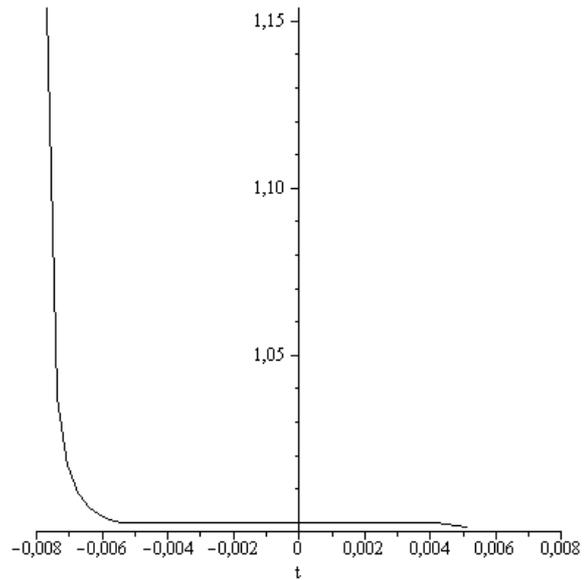

Fig. 47. v – potential $\tilde{\varphi}$. (the second approximation, Variant 6).

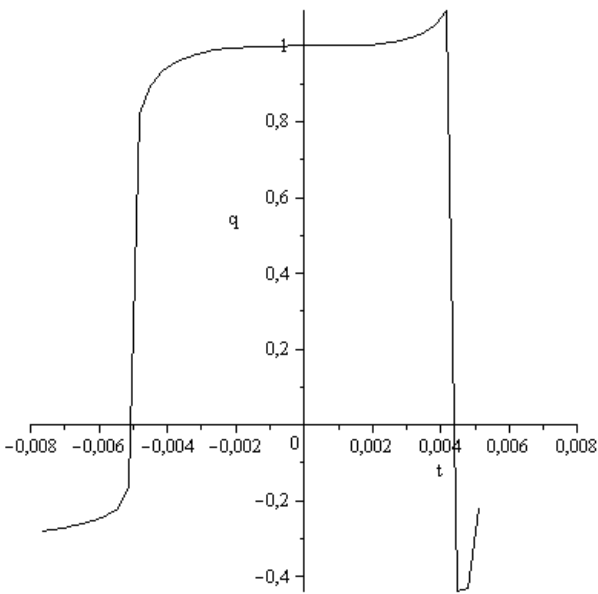

Fig. 48. q – the negative particles pressure. (the second approximation, Variant 6).

## 5. Results of the mathematical modeling with the external electric field.

Let us consider now the results of the mathematical modeling with taking into account the intensity of the external electric field which does not depend on $y$. In this case the solution of the hydrodynamic system (3) – (8) should be found. After averaging and in the moving coordinate system it leads to the following equations written in the first approximation (compare with the system (27) – (32)):

Dimensionless Poisson equation for the self-consistent electric field:

$$\frac{\partial^2 \tilde{\varphi}}{\partial \tilde{\xi}^2} = -4\pi R \left\{ \frac{m_e}{m_p} \left[ \tilde{\rho}_p - \frac{m_e H}{m_p \tilde{u}^2} \frac{\partial}{\partial \tilde{\xi}} (\tilde{\rho}_p (\tilde{u}-1)) \right] - \left[ \tilde{\rho}_e - \frac{H}{\tilde{u}^2} \frac{\partial}{\partial \tilde{\xi}} (\tilde{\rho}_e (\tilde{u}-1)) \right] \right\}. \tag{33}$$

Continuity equation for the positive particles:

$$\frac{\partial}{\partial \tilde{\xi}} [\tilde{\rho}_p (1-\tilde{u})] + \frac{m_e}{m_p} \frac{\partial}{\partial \tilde{\xi}} \left\{ \frac{H}{\tilde{u}^2} \frac{\partial}{\partial \tilde{\xi}} [\tilde{\rho}_p (\tilde{u}-1)^2] \right\} + \frac{m_e}{m_p} \frac{\partial}{\partial \tilde{\xi}} \left\{ \frac{H}{\tilde{u}^2} \left[ \frac{V_{0p}^2}{u_0^2} \frac{\partial}{\partial \tilde{\xi}} \tilde{p}_p - \right.\right.$$
$$\left.\left. - \frac{m_e}{m_p} \tilde{\rho}_p E \left( -\frac{\partial \tilde{\varphi}}{\partial \tilde{\xi}} + \tilde{U}'_{11} \sin\left(\frac{4\pi}{3\tilde{a}} \tilde{\xi} - \frac{\pi}{3}\right) + \tilde{E}_0 \right) \right] \right\} = 0 \tag{34}$$

Continuity equation for electrons:

$$\frac{\partial}{\partial \tilde{\xi}} [\tilde{\rho}_e (1-\tilde{u})] + \frac{\partial}{\partial \tilde{\xi}} \left\{ \frac{H}{\tilde{u}^2} \frac{\partial}{\partial \tilde{\xi}} [\tilde{\rho}_e (\tilde{u}-1)^2] \right\} + \frac{\partial}{\partial \tilde{\xi}} \left\{ \frac{H}{\tilde{u}^2} \left[ \frac{V_{0e}^2}{u_0^2} \frac{\partial}{\partial \tilde{\xi}} \tilde{p}_e - \right.\right.$$
$$\left.\left. - \tilde{\rho}_e E \left( \frac{\partial \tilde{\varphi}}{\partial \tilde{\xi}} - \tilde{U}'_{11} \sin\left(\frac{4\pi}{3\tilde{a}} \tilde{\xi} - \frac{\pi}{3}\right) - \tilde{E}_0 \right) \right] \right\} = 0 \tag{35}$$

Momentum equation for the $x$ direction:

$$\frac{\partial}{\partial \tilde{\xi}} \left\{ (\tilde{\rho}_p + \tilde{\rho}_e) \tilde{u} (\tilde{u}-1) + \frac{V_{0p}^2}{u_0^2} \tilde{p}_p + \frac{V_{0e}^2}{u_0^2} \tilde{p}_e \right\} -$$
$$- \frac{m_e}{m_p} \tilde{\rho}_p E \left( -\frac{\partial \tilde{\varphi}}{\partial \tilde{\xi}} + \tilde{U}'_{11} \sin\left(\frac{4\pi}{3\tilde{a}} \tilde{\xi} - \frac{\pi}{3}\right) + \tilde{E}_0 \right) -$$
$$- \tilde{\rho}_e E \left( \frac{\partial \tilde{\varphi}}{\partial \tilde{\xi}} - \tilde{U}'_{11} \sin\left(\frac{4\pi}{3\tilde{a}} \tilde{\xi} - \frac{\pi}{3}\right) - \tilde{E}_0 \right) +$$
$$+ \frac{m_e}{m_p} \frac{\partial}{\partial \tilde{\xi}} \left\{ \frac{H}{\tilde{u}^2} \left[ \frac{\partial}{\partial \tilde{\xi}} \left( 2 \frac{V_{0p}^2}{u_0^2} \tilde{p}_p (1-\tilde{u}) - \tilde{\rho}_p \tilde{u} (1-\tilde{u})^2 \right) - \right.\right.$$
$$\left.\left. - \frac{m_e}{m_p} \tilde{\rho}_p (1-\tilde{u}) E \left( -\frac{\partial \tilde{\varphi}}{\partial \tilde{\xi}} + \tilde{U}'_{11} \sin\left(\frac{4\pi}{3\tilde{a}} \tilde{\xi} - \frac{\pi}{3}\right) + \tilde{E}_0 \right) \right] \right\} +$$
$$+ \frac{\partial}{\partial \tilde{\xi}} \left\{ \frac{H}{\tilde{u}^2} \left[ \frac{\partial}{\partial \tilde{\xi}} \left( 2 \frac{V_{0e}^2}{u_0^2} \tilde{p}_e (1-\tilde{u}) - \tilde{\rho}_e \tilde{u} (1-\tilde{u})^2 \right) - \tilde{\rho}_e (1-\tilde{u}) E \left( \frac{\partial \tilde{\varphi}}{\partial \tilde{\xi}} - \tilde{U}'_{11} \sin\left(\frac{4\pi}{3\tilde{a}} \tilde{\xi} - \frac{\pi}{3}\right) - \tilde{E}_0 \right) \right] \right\} +$$
$$+ \frac{H}{\tilde{u}^2} E \left( \frac{m_e}{m_p} \right)^2 \left( -\frac{\partial \tilde{\varphi}}{\partial \tilde{\xi}} + \tilde{U}'_{11} \sin\left(\frac{4\pi}{3\tilde{a}} \tilde{\xi} - \frac{\pi}{3}\right) + \tilde{E}_0 \right) \left( \frac{\partial}{\partial \tilde{\xi}} (\tilde{\rho}_p (\tilde{u}-1)) \right) +$$
$$+ \frac{H}{\tilde{u}^2} E \left( \frac{\partial \tilde{\varphi}}{\partial \tilde{\xi}} - \tilde{U}'_{11} \sin\left(\frac{4\pi}{3\tilde{a}} \tilde{\xi} - \frac{\pi}{3}\right) - \tilde{E}_0 \right) \left( \frac{\partial}{\partial \tilde{\xi}} (\tilde{\rho}_e (\tilde{u}-1)) \right) -$$

$$-\frac{m_e}{m_p}\frac{\partial}{\partial \tilde{\xi}}\left\{\frac{H}{\tilde{u}^2}\frac{V_{0p}^2}{u_0^2}\frac{\partial}{\partial \tilde{\xi}}(\tilde{p}_p\tilde{u})\right\}-\frac{\partial}{\partial \tilde{\xi}}\left\{\frac{H}{\tilde{u}^2}\frac{V_{0e}^2}{u_0^2}\frac{\partial}{\partial \tilde{\xi}}(\tilde{p}_e\tilde{u})\right\}+$$

$$+\left(\frac{m_e}{m_p}\right)^2 E\frac{\partial}{\partial \tilde{\xi}}\left\{\frac{H}{\tilde{u}^2}\left[\left(-\frac{\partial \tilde{\varphi}}{\partial \tilde{\xi}}+\tilde{U}'_{11}\sin\left(\frac{4\pi}{3\tilde{a}}\tilde{\xi}-\frac{\pi}{3}\right)+\tilde{E}_0\right)\tilde{\rho}_p\tilde{u}\right]\right\}+ \quad (36)$$

$$+E\frac{\partial}{\partial \tilde{\xi}}\left\{\frac{H}{\tilde{u}^2}\left[\left(\frac{\partial \tilde{\varphi}}{\partial \tilde{\xi}}-\tilde{U}'_{11}\sin\left(\frac{4\pi}{3\tilde{a}}\tilde{\xi}-\frac{\pi}{3}\right)-\tilde{E}_0\right)\tilde{\rho}_e\tilde{u}\right]\right\}=0$$

Energy equation for the positive particles:

$$\frac{\partial}{\partial \tilde{\xi}}\left[\tilde{\rho}_p\tilde{u}^2(\tilde{u}-1)+5\frac{V_{0p}^2}{u_0^2}\tilde{p}_p\tilde{u}-3\frac{V_{0p}^2}{u_0^2}\tilde{p}_p\right]-2\frac{m_e}{m_p}\tilde{\rho}_p E\left(-\frac{\partial \tilde{\varphi}}{\partial \tilde{\xi}}+\tilde{U}'_{11}\sin\left(\frac{4\pi}{3\tilde{a}}\tilde{\xi}-\frac{\pi}{3}\right)+\tilde{E}_0\right)\tilde{u}+$$

$$+\frac{\partial}{\partial \tilde{\xi}}\left\{\frac{H}{\tilde{u}^2}\frac{m_e}{m_p}\left[\frac{\partial}{\partial \tilde{\xi}}\left(-\tilde{\rho}_p\tilde{u}^2(1-\tilde{u})^2+7\frac{V_{0p}^2}{u_0^2}\tilde{p}_p\tilde{u}(1-\tilde{u})+3\frac{V_{0p}^2}{u_0^2}\tilde{p}_p(\tilde{u}-1)-\frac{V_{0p}^2}{u_0^2}\tilde{p}_p\tilde{u}^2-5\frac{V_{0p}^4}{u_0^4}\frac{\tilde{p}_p^2}{\tilde{\rho}_p}\right)+\right.$$

$$+E\left(-2\frac{m_e}{m_p}\tilde{\rho}_p\tilde{u}(1-\tilde{u})+\frac{m_e}{m_p}\tilde{\rho}_p\tilde{u}^2+5\frac{m_e}{m_p}\frac{V_{0p}^2}{u_0^2}\tilde{p}_p\right)\left(-\frac{\partial \tilde{\varphi}}{\partial \tilde{\xi}}+\right.$$

$$\left.\left.+\tilde{U}'_{11}\sin\left(\frac{4\pi}{3\tilde{a}}\tilde{\xi}-\frac{\pi}{3}\right)+\tilde{E}_0\right)\right]\right\}+2\frac{H}{\tilde{u}^2}E\left(\frac{m_e}{m_p}\right)^2\left[-\frac{\partial}{\partial \tilde{\xi}}(\tilde{\rho}_p\tilde{u}(1-\tilde{u}))+\right.$$

$$\left.+\frac{V_{0p}^2}{u_0^2}\frac{\partial}{\partial \tilde{\xi}}\tilde{p}_p\right]\left(-\frac{\partial \tilde{\varphi}}{\partial \tilde{\xi}}+\tilde{U}'_{11}\sin\left(\frac{4\pi}{3\tilde{a}}\tilde{\xi}-\frac{\pi}{3}\right)+\tilde{E}_0\right)-$$

$$-2\frac{H}{\tilde{u}^2}E^2\left(\frac{m_e}{m_p}\right)^3\tilde{\rho}_p\left[\left(-\frac{\partial \tilde{\varphi}}{\partial \tilde{\xi}}+\tilde{U}'_{11}\sin\left(\frac{4\pi}{3\tilde{a}}\tilde{\xi}-\frac{\pi}{3}\right)+\tilde{E}_0\right)^2+\frac{1}{2}\left(\tilde{U}'_{10}\sin\left(\frac{2\pi}{3\tilde{a}}\tilde{\xi}+\frac{\pi}{3}\right)\right)^2+\right.$$

$$\left.+\frac{3}{2}\left(\tilde{U}'_{10}\cos\left(\frac{2\pi}{3\tilde{a}}\tilde{\xi}+\frac{\pi}{3}\right)\right)^2+6(\tilde{U}'_{11})^2+\frac{16}{\pi}(\tilde{U}'_{10}\tilde{U}'_{11})\cos\left(\frac{2\pi}{3\tilde{a}}\tilde{\xi}+\frac{\pi}{3}\right)\right]=$$

$$=-\frac{\tilde{u}^2}{Hu_0^2}(V_{0p}^2\tilde{p}_p-\tilde{p}_e V_{0e}^2)\left(1+\frac{m_p}{m_e}\right) \quad (37)$$

Energy equation for electrons:

$$\frac{\partial}{\partial \tilde{\xi}}\left[\tilde{\rho}_e\tilde{u}^2(\tilde{u}-1)+5\frac{V_{0e}^2}{u_0^2}\tilde{p}_e\tilde{u}-3\frac{V_{0e}^2}{u_0^2}\tilde{p}_e\right]-2\tilde{\rho}_e\tilde{u}E\left(\frac{\partial \tilde{\varphi}}{\partial \tilde{\xi}}-\tilde{U}'_{11}\sin\left(\frac{4\pi}{3\tilde{a}}\tilde{\xi}-\frac{\pi}{3}\right)-\tilde{E}_0\right)+$$

$$+\frac{\partial}{\partial \tilde{\xi}}\left\{\frac{H}{\tilde{u}^2}\left[\frac{\partial}{\partial \tilde{\xi}}\left(-\tilde{\rho}_e\tilde{u}^2(1-\tilde{u})^2+7\frac{V_{0e}^2}{u_0^2}\tilde{p}_e\tilde{u}(1-\tilde{u})+3\frac{V_{0e}^2}{u_0^2}\tilde{p}_e(\tilde{u}-1)-\frac{V_{0e}^2}{u_0^2}\tilde{p}_e\tilde{u}^2-5\frac{V_{0e}^4}{u_0^4}\frac{\tilde{p}_e^2}{\tilde{\rho}_e}\right)+\right.$$

$$\left.\left.+E\left(-2\tilde{\rho}_e\tilde{u}(1-\tilde{u})+\tilde{\rho}_e\tilde{u}^2+5\frac{V_{0e}^2}{u_0^2}\tilde{p}_e\right)\left(\frac{\partial \tilde{\varphi}}{\partial \tilde{\xi}}-\tilde{U}'_{11}\sin\left(\frac{4\pi}{3\tilde{a}}\tilde{\xi}-\frac{\pi}{3}\right)-\tilde{E}_0\right)\right]\right\}+$$

$$+E\left(-2\frac{H}{\tilde{u}^2}\frac{\partial}{\partial \tilde{\xi}}(\tilde{\rho}_e\tilde{u}(1-\tilde{u}))+2\frac{H}{\tilde{u}^2}\frac{V_{0e}^2}{u_0^2}\frac{\partial}{\partial \tilde{\xi}}\tilde{p}_e\right)\left(\frac{\partial \tilde{\varphi}}{\partial \tilde{\xi}}-\tilde{U}'_{11}\sin\left(\frac{4\pi}{3\tilde{a}}\tilde{\xi}-\frac{\pi}{3}\right)-\tilde{E}_0\right)-$$

$$-2E^2 \frac{H}{\tilde{u}^2}\tilde{\rho}_e\left[\left(-\frac{\partial\tilde{\varphi}}{\partial\tilde{\xi}}+\tilde{U}'_{11}\sin\left(\frac{4\pi}{3\tilde{a}}\tilde{\xi}-\frac{\pi}{3}\right)+\tilde{E}_0\right)^2+\right.$$

$$\left.+\frac{1}{2}\left(\tilde{U}'_{10}\sin\left(\frac{2\pi}{3\tilde{a}}\tilde{\xi}+\frac{\pi}{3}\right)\right)^2+\frac{3}{2}\left(\tilde{U}'_{10}\cos\left(\frac{2\pi}{3\tilde{a}}\tilde{\xi}+\frac{\pi}{3}\right)\right)^2+6(\tilde{U}'_{11})^2+\frac{16}{\pi}(\tilde{U}'_{10}\tilde{U}'_{11})\cos\left(\frac{2\pi}{3\tilde{a}}\tilde{\xi}+\frac{\pi}{3}\right)\right]=$$

$$=-\frac{\tilde{u}^2}{Hu_0^2}\left(V_{0e}^2\tilde{p}_e-V_{0p}^2\tilde{p}_p\right)\left(1+\frac{m_p}{m_e}\right) \qquad (38)$$

Two classes of parameters were used by the mathematical modeling – parameters and scales which were not changed during calculations and varied parameters indicated in Table 4.

Parameters, scales and Cauchy conditions which are common for modeling with the external field:

$\frac{m_e}{m_p}=5\cdot 10^{-5}$, the scales $\rho_0=10^{-10}\,g/cm^3$, $u_0=5\cdot 10^6\,cm/s$, $V_{0e}=5\cdot 10^6\,cm/s$, $V_{0p}=5\cdot 10^4\,cm/s$, $x_0=a=0.142\,nm$, $\varphi_0=10^{-4}\frac{e}{a}=3.4\cdot 10^{-6}\,CGSE_\varphi$.

Dimensionless parameters $R=3\cdot 10^{-3}$, $E=0.1$, $H=15$ (by $N_R=1$). Admit that for the lattice $U\sim V_{1,(10)}\sim V_{1,(11)}\sim\varphi_0$ and choose $\tilde{U}'_{10}=10$, $\tilde{U}'_{11}=10$.

Cauchy conditions $\tilde{\rho}_e(0)=1$, $\tilde{\rho}_p(0)=2\cdot 10^4$, $\tilde{p}_e(0)=1$, $\tilde{p}_p(0)=2\cdot 10^4$, $\tilde{\varphi}(0)=1$, $\frac{\partial\tilde{\rho}_e}{\partial\tilde{\xi}}(0)=0$, $\frac{\partial\tilde{\rho}_p}{\partial\tilde{\xi}}(0)=0$.

Table 4. Varied parameters in calculations with the external electric field.

| Variant № | $\tilde{E}_0$ | $\frac{\partial\tilde{\varphi}}{\partial\tilde{\xi}}(0)$ | $\frac{\partial\tilde{p}_p}{\partial\tilde{\xi}}(0)$ | $\frac{\partial\tilde{p}_e}{\partial\tilde{\xi}}(0)$ |
|---|---|---|---|---|
| 1 | 0 | 0 | 0 | 0 |
| 7.0 | 10 | 10 | 0 | 0 |
| 7.1 | 10 | 10 | 10 | -1 |
| 8.0 | 100 | 100 | 0 | 0 |
| 8.1 | 100 | 100 | 10 | 0 |
| 9.0 | 10000 | 10000 | 0 | 0 |
| 9.1 | 10000 | 10000 | 10 | -1 |

The external intensity of the electric field is written as $E_0 = \frac{\varphi_0}{x_0}\tilde{E}_0 = 10^{-4}\frac{e}{a^2}\tilde{E}_0 = 238 СГСЭ_E \tilde{E}_0 = 7.14\cdot 10^6 \frac{B}{м}\tilde{E}_0$. It means that even by $\tilde{E}_0 = 1$ we are dealing with the rather strong fields. But namely strong external fields can exert the influence on the soliton structures compared with the Coulomb forces in the lattice. For example in [20] the influence of the external electric field in graphene up to $10^7 - 10^8 V/m$. The values $\tilde{E}_0$ are indicated in Table 4, variants 9.0 and 9.1 respond to the extremely strong external field.

Table 4 contains in the first line the reminder about the first variant of calculations reflected on figures 2 – 5. These data (in the absence of the external field, $\tilde{E}_0 = 0$) are convenient for the following result comparison. The variants of calculations in Table 4 are grouped on principle of the $\tilde{E}_0$ increasing. In more details: figures 49 – 58 correspond to $\tilde{E}_0 = 10$, figures 59 – 68 correspond to $\tilde{E}_0 = 100$, figures 69 – 80 correspond to $\tilde{E}_0 = 10000$.

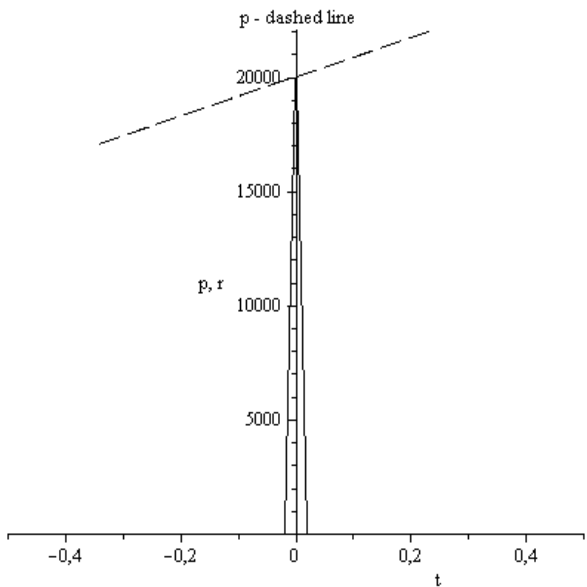 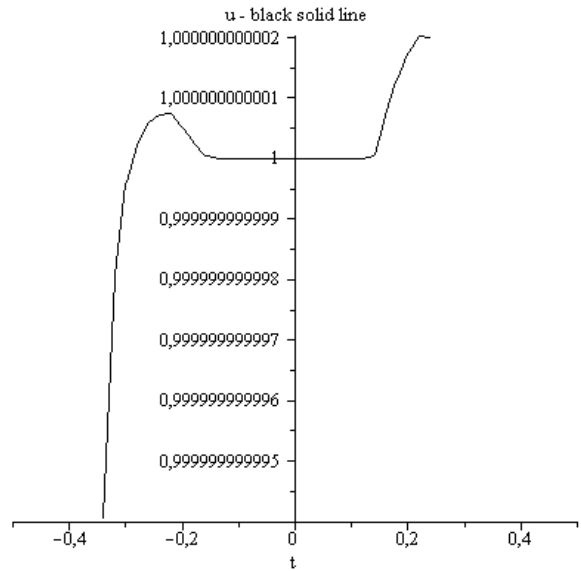

Fig. 49. r – the positive particles density, (solid line); p – the positive particles pressure. (Variant 7.0).

Fig. 50. u – velocity $\tilde{u}$. (Variant 7.0).

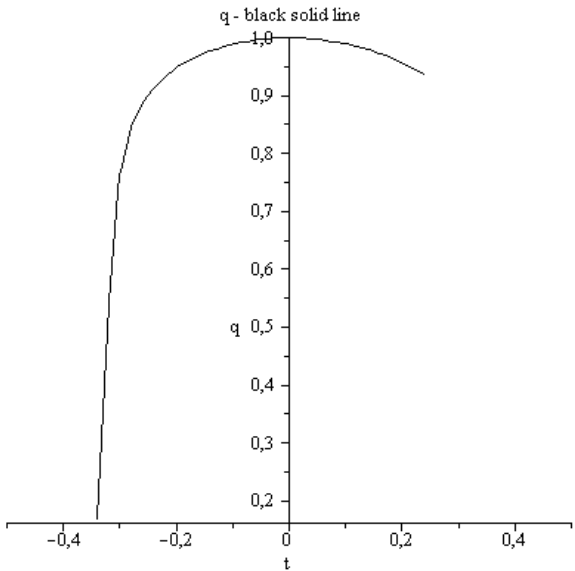

Fig. 51. q – the negative particles pressure. (Variant 7.0).

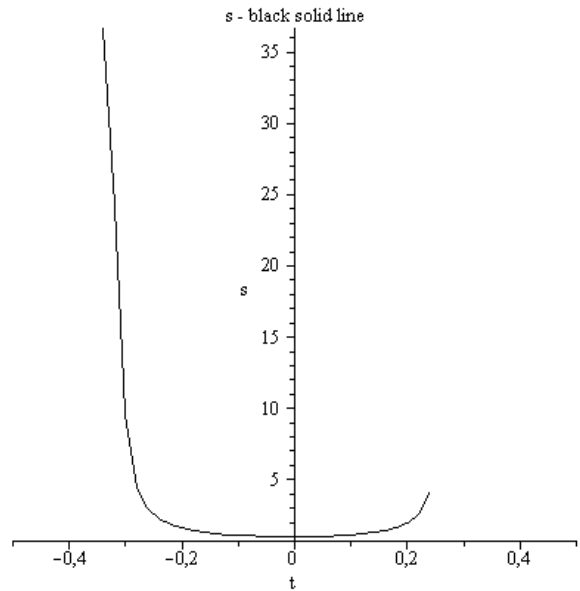

Fig. 52. s – electron density $\tilde{\rho}_e$, (Variant 7.0).

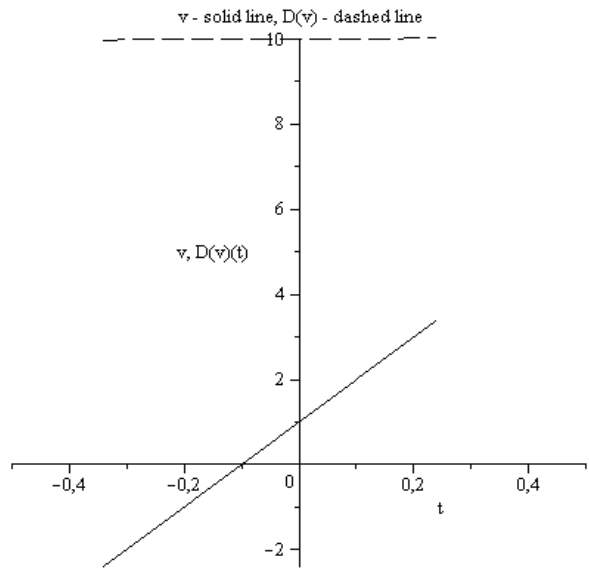

Fig. 53. v – potential $\tilde{\varphi}$ (solid line); $D(v)(t)$, (Variant 7.0).

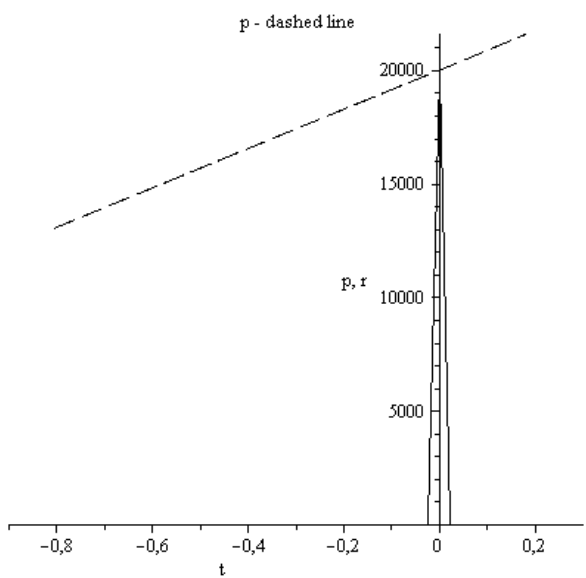
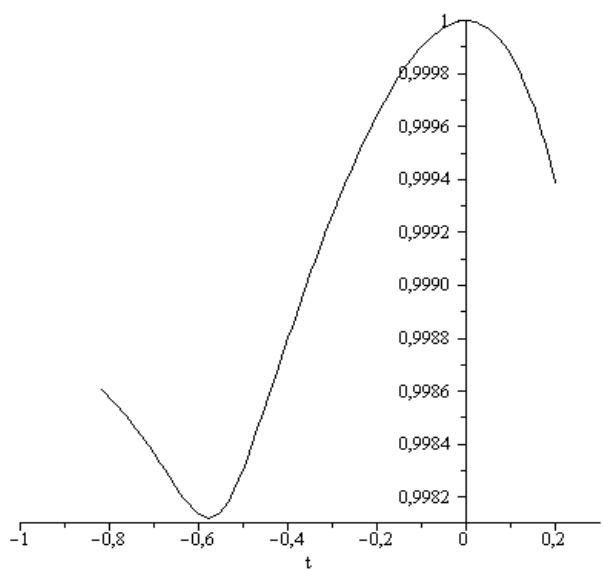

Fig. 54. r – the positive particles density, (solid line); p – the positive particles pressure. (Variant 7.1).

Fig. 55. u – velocity $\tilde{u}$. (Variant 7.1).

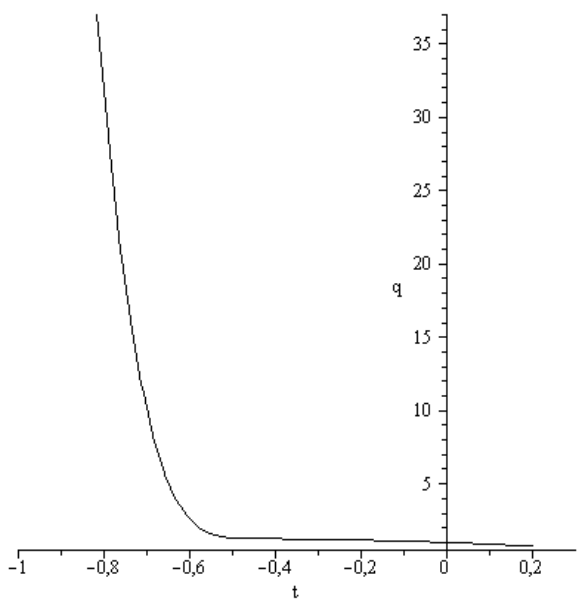
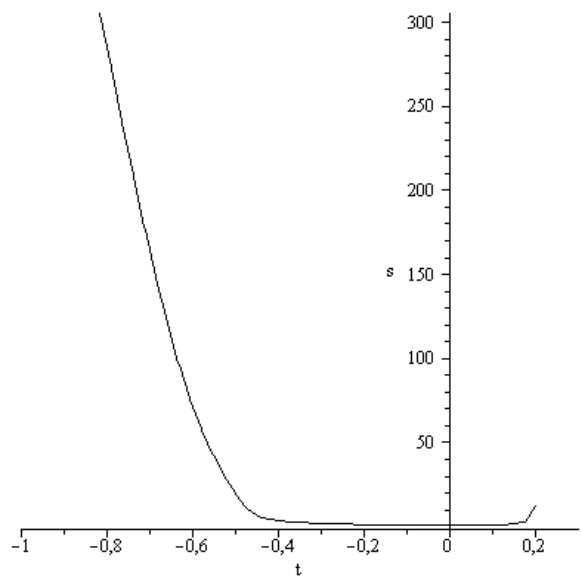

Fig. 56. q – the negative particles pressure. (Variant 7.1).

Fig. 57. s – electron density $\tilde{\rho}_e$, (Variant 7.1).

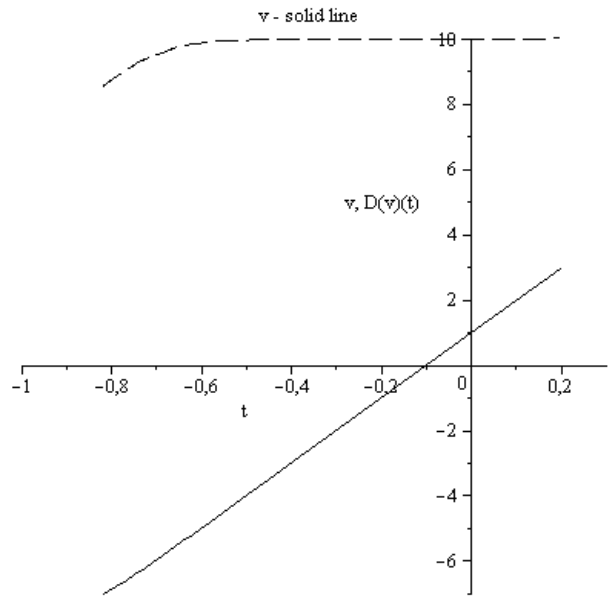

Fig. 58. v – potential $\widetilde{\varphi}$ (solid line);

$D(v)(t)$, (Variant 7.1).

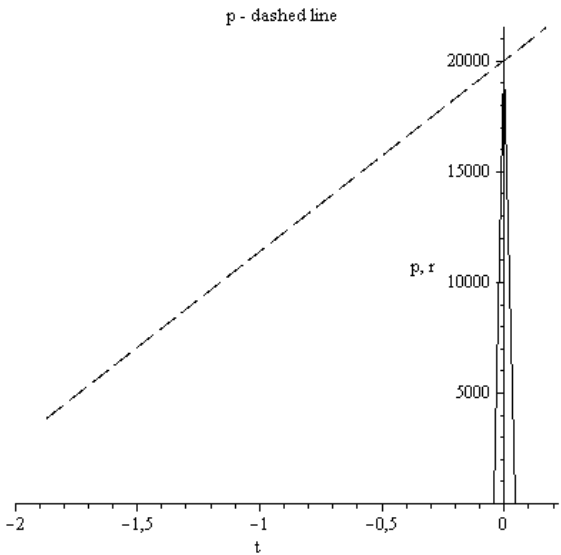

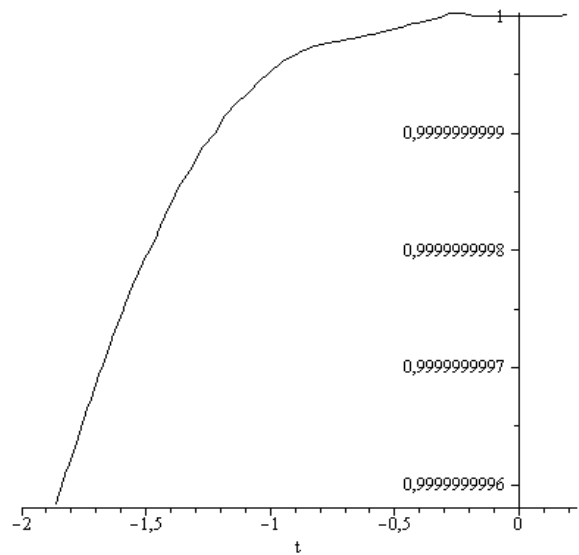

Fig. 59. r – the positive particles density, (solid line); p – the positive particles pressure. (Variant 8.0).

Fig. 60. u – velocity $\widetilde{u}$. (Variant 8.0).

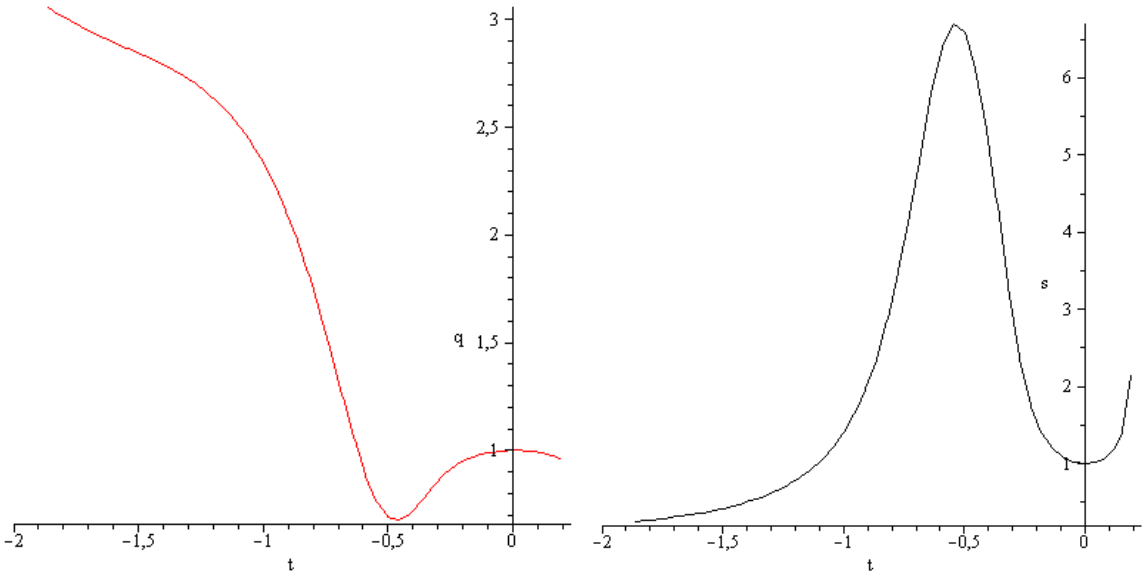

Fig. 61. q – the negative particles pressure. (Variant 8.0).

Fig. 62. s – electron density $\tilde{\rho}_e$, (Variant 8.0).

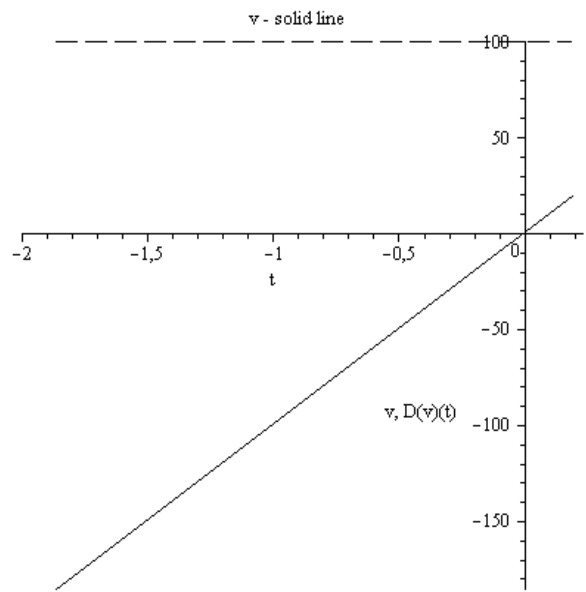

Fig. 63. v – potential $\tilde{\varphi}$ (solid line); $D(v)(t)$, (Variant 8.0).

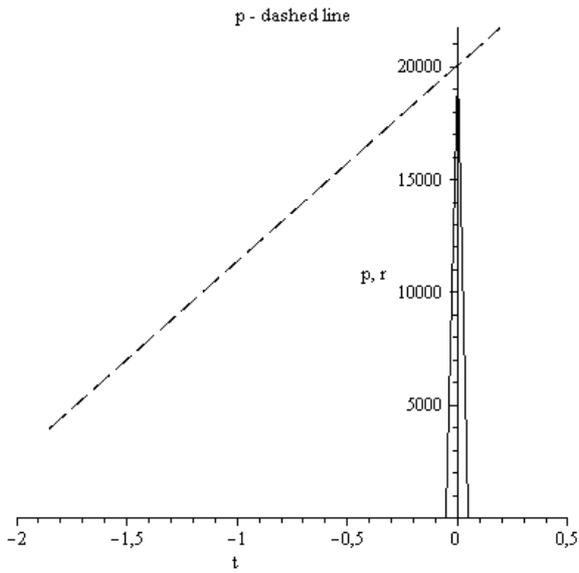

Fig. 64. r – the positive particles density, (solid line); p – the positive particles pressure. (Variant 8.1).

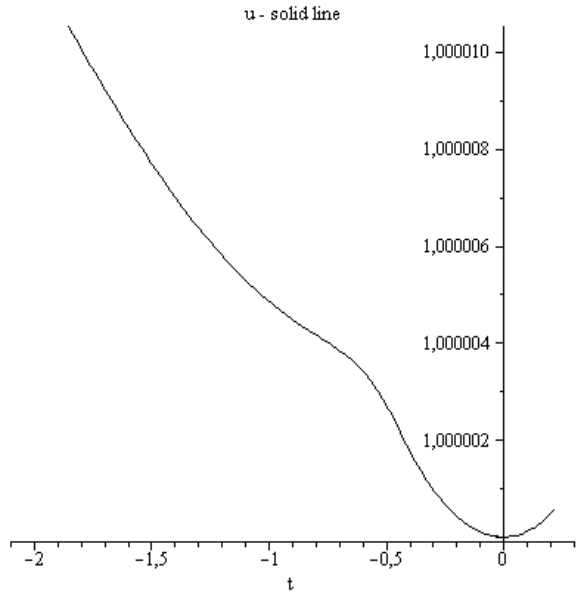

Fig. 65. u – velocity $\tilde{u}$. (Variant 8.1).

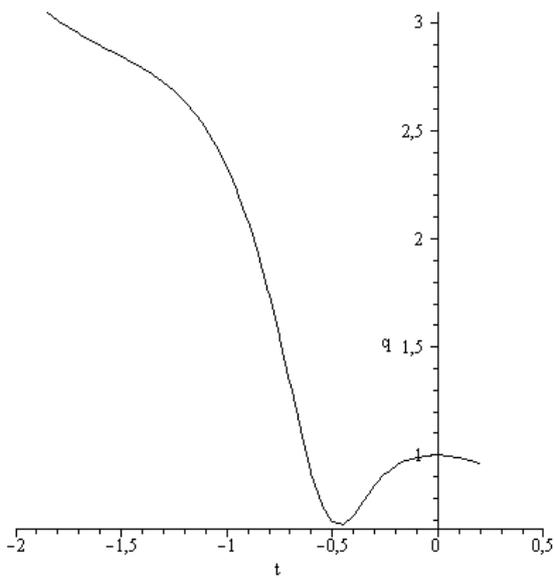

Fig. 66. q – the negative particles pressure. (Variant 8.1).

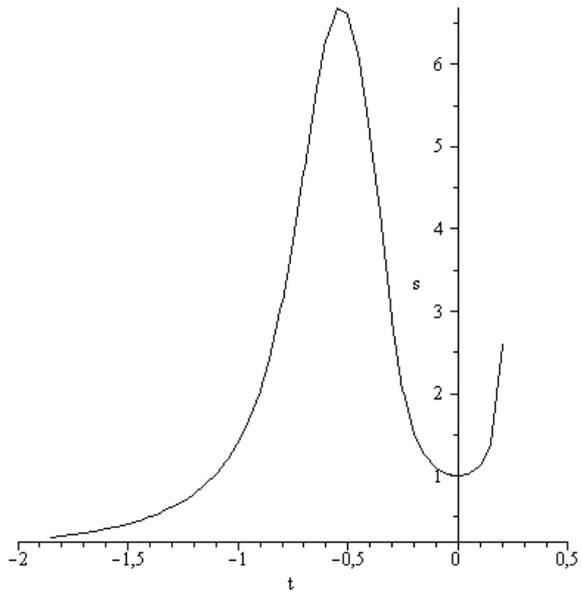

Fig. 67. s – electron density $\tilde{\rho}_e$, (Variant 8.1).

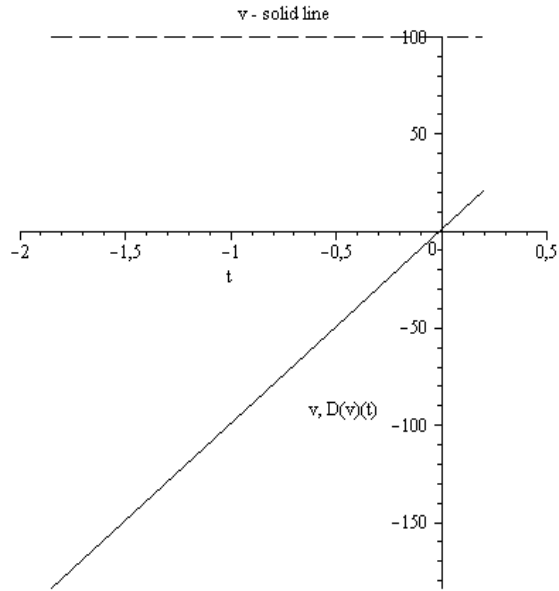

Fig. 68. v – potential $\tilde{\varphi}$ (solid line);

$D(v)(t)$, (Variant 8.1).

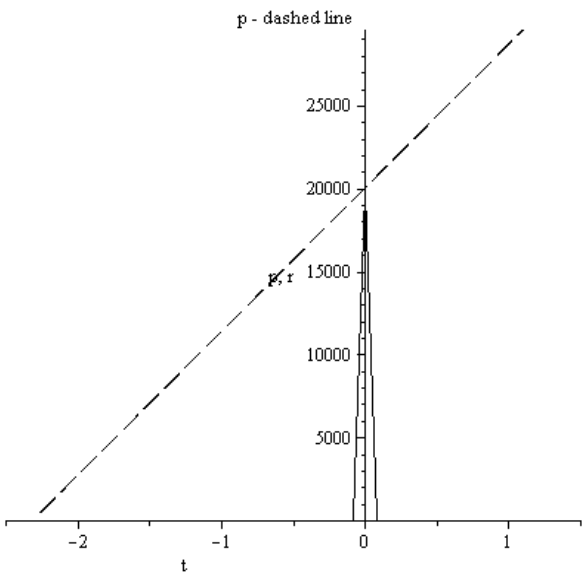

Fig. 69. r – the positive particles density,
(solid line); p – the positive particles pressure.
(Variant 9.0).

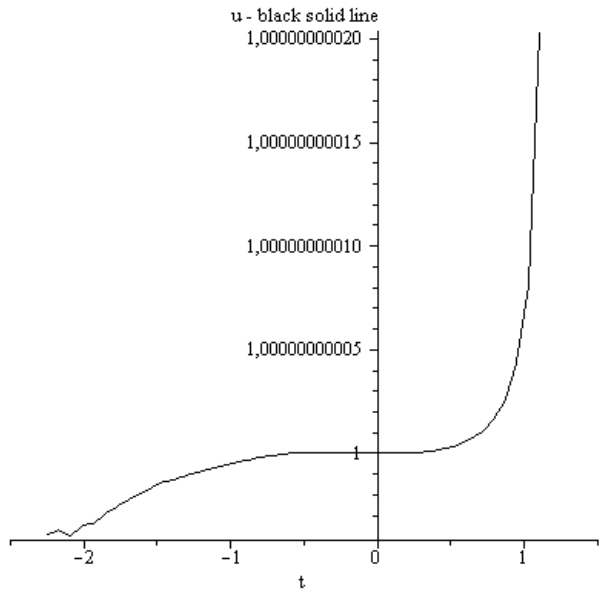

Fig. 70. u – velocity $\tilde{u}$. (Variant 9.0).

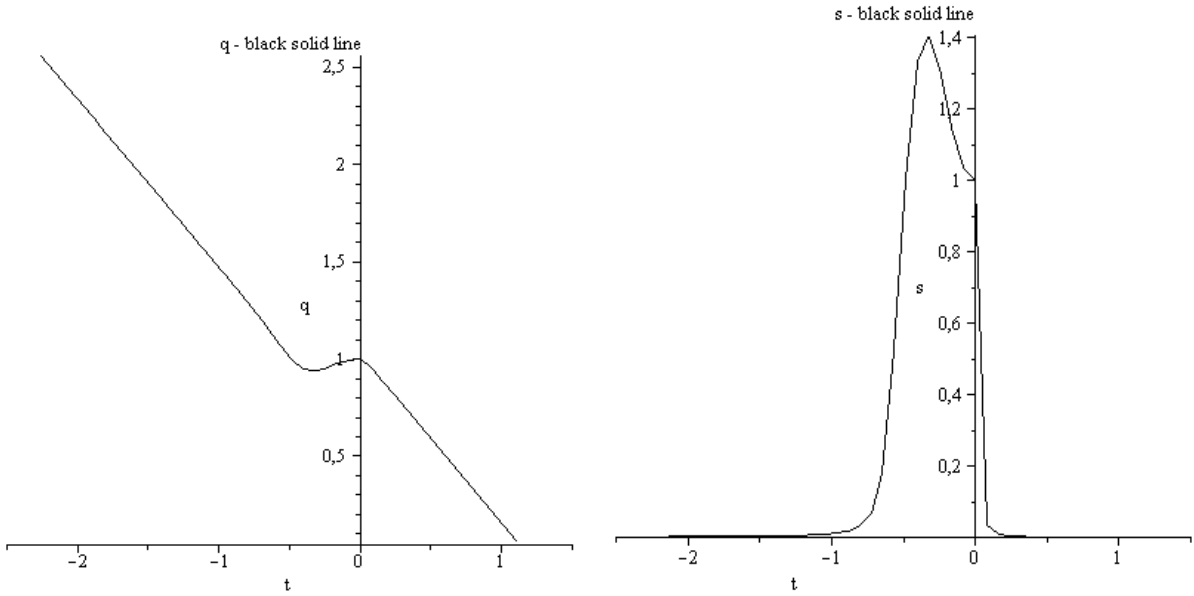

Fig. 71. q – the negative particles pressure. (Variant 9.0).    Fig. 72. s – electron density $\tilde{\rho}_e$, (Variant 9.0).

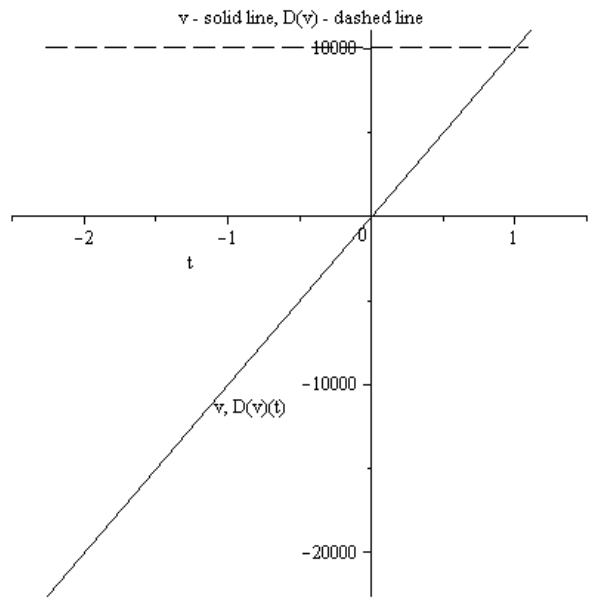

Fig. 73. v – potential $\tilde{\varphi}$ (solid line); $D(v)(t)$, (Variant 9.0).

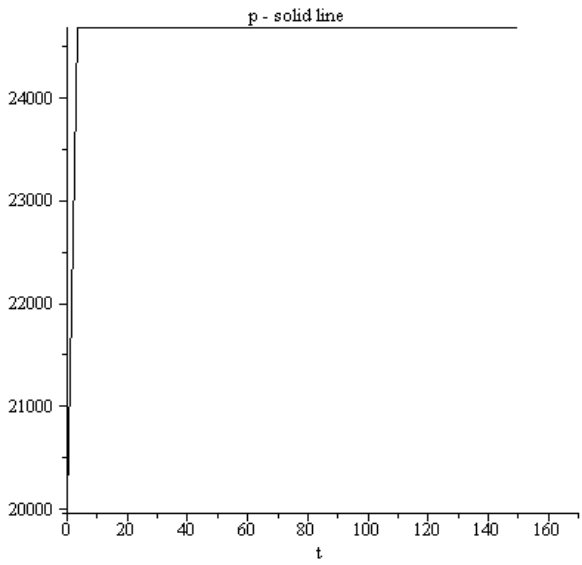

Fig. 74. p – the positive particles pressure. (Variant 9.1).

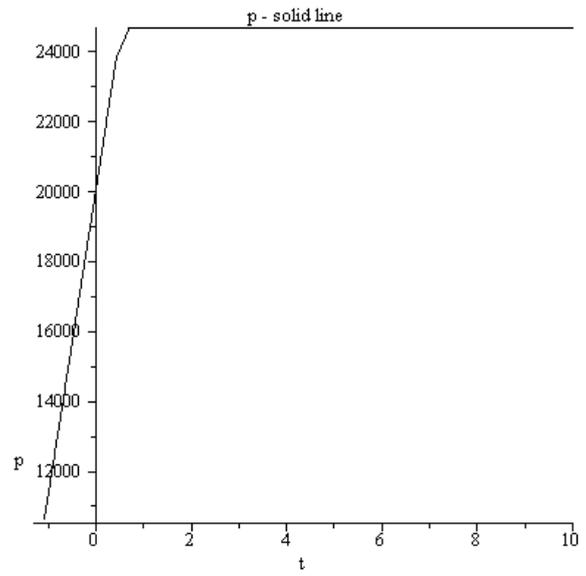

Fig. 75. p – the positive particles pressure. (Variant 9.1).

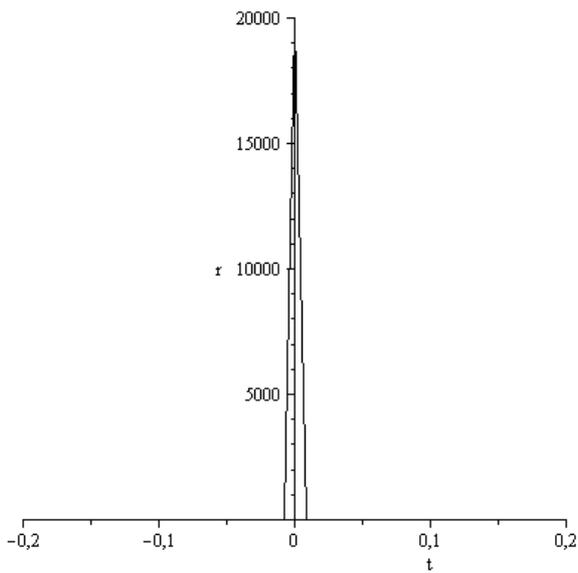

Fig. 76. r – the positive particles density, (Variant 9.1).

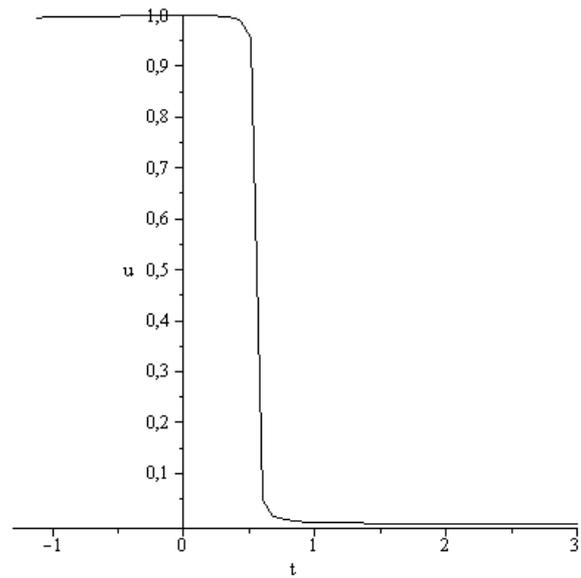

Fig. 77. u – velocity $\tilde{u}$. (Variant 9.1).

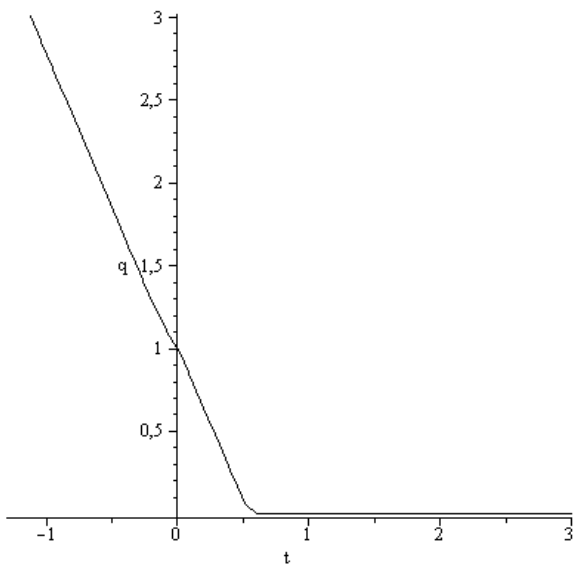
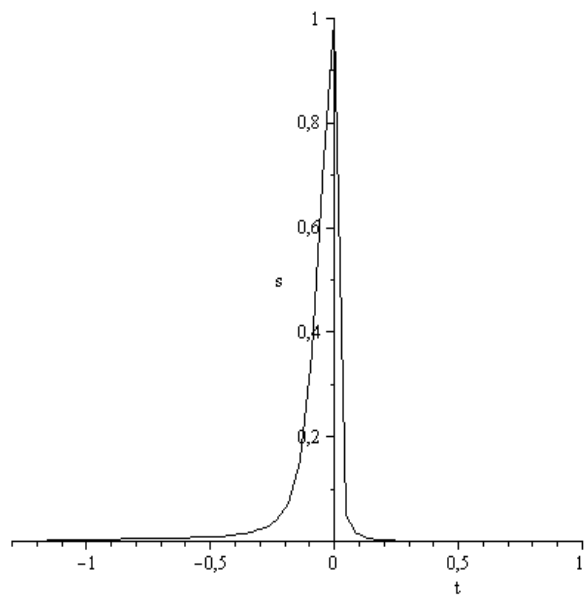

Fig. 78. q – the negative particles pressure. (Variant 9.1).

Fig. 79. s – electron density $\tilde{\rho}_e$, (Variant 9.1).

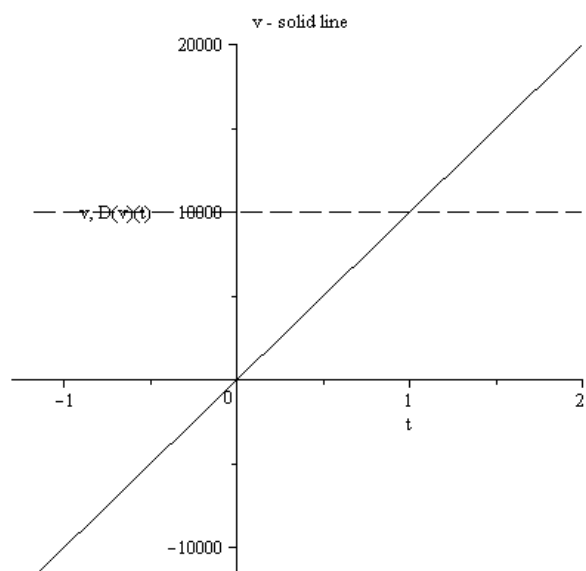

Fig. 80. v – potential $\tilde{\varphi}$ (solid line); $D(v)(t)$, (Variant 9.1).

Consider now the character features of the soliton evolution and the change of the charge distribution in solitons with growing of the external field intensity:
1. The character soliton size is defined by the area where $\tilde{u} = 1$. It means that all part of the soliton wave are moving without destruction. The size of this area is practically independent on the choose of the numerical method of calculations.

2. Figures 75 – 77 demonstrate the typical situation when the area of possible numerical calculations for a physical variable does not coincide with area $\tilde{u} = 1$ where the soliton regime exists.

3. In the area of the soliton existence the condition $\tilde{u} = 1$ is fulfilled with the high accuracy defined practically by accuracy of the choosed numerical method (see Figs. 50, 55, 60, 65, 70, 77).

4. As a rule for the choosed topology of the electric field the size of the soliton existence is growing with increasing of the electric field intensity.

5. Under the influence of the external electric field the captured electron cloud is displacing in the opposite direction (of the negative variable $\tilde{\xi}$). The soliton kernel is loosing its symmetry.

6. The redistribution of the self-consistent effective charge creates the self-consistence field with the opposite (to the external field) direction, (see Figs. 53, 58, 63, 68, 73, 80).

7. The quantum pressure of the positive particle is growing with the $\tilde{\xi}$ increase. On the whole the specific features of the $\tilde{p}$, $\tilde{q}$ pressures are defined by the process of the soliton formation.

**Conclusion**

       The origin of the charge density waves (CDW) is a long-standing problem relevant to a number of important issues in condensed matter physics. Mathematical modeling of the CDW expansion as well as the problem of the high temperature superconduction can be solved only on the basement of the nonlocal quantum hydrodynamics in particular on the basement of the Alexeev non-local quantum hydrodynamics. It is known that the Schrödinger – Madelung quantum physics leads to the destruction of the wave packets and can not be used for the solution of this kind of problems. The appearance in mathematics the soliton solutions is the rare and remarkable effect. As we see the soliton's appearance in the generalized hydrodynamics created by Alexeev is an "ordinary" oft-recurring fact. The realized here mathematical modeling CDW expansion support established in [1, 3] mechanism of the relay ("estafette") motion of the soliton' system ("lattice ion – electron") which is realizing by the absence of chemical bonds. Important to underline that the soliton mechanism of CDW expansion in graphene (and other substances like $NbSe_2$) takes place in the extremely large diapason of physical parameters. But CDW existence belongs to effects convoying the high temperature superconductivity. It means that the high temperature superconductivity can be explained in the frame of the non-local soliton quantum hydrodynamics.